\documentclass[twocolumn]{revtex4-2}

\usepackage[utf8]{inputenc}
\usepackage[T1]{fontenc}
\usepackage[svgnames, dvipsnames]{xcolor}
\usepackage{dcolumn}
\usepackage{amsfonts, stmaryrd, mathtools, mathrsfs, abraces, bm}
\usepackage{bbm}
\DeclareMathAlphabet{\mathpzc}{OT1}{pzc}{m}{it}
\usepackage{graphicx}
\usepackage[toc,page]{appendix}
\usepackage[colorlinks=true, linkcolor=MediumBlue, citecolor=blue,
urlcolor=blue,breaklinks=true]{hyperref}
\usepackage{mathtools}
\usepackage{xspace}
\newcommand\1{\mathbbm{1}}
\usepackage[boxed,noline]{algorithm2e}
\renewcommand{\d}{\mathrm{d}}
\newcommand*{\setind}{\llbracket 1, N \rrbracket}
\newcommand\mydots{\hbox to 0.7em{.\hss.\hss.}\thinspace}

\begin{document}

\title[Necessary and sufficient symmetries in Event-Chain Monte Carlo with generalized flows]{Necessary and sufficient symmetries in Event-Chain Monte Carlo with
  generalized flows and application to hard dimers}

 \author{Tristan Guyon} \email{tristan.guyon@cnrs.fr}
\author{Arnaud Guillin} \email{arnaud.guillin@uca.fr} \author{Manon Michel}
\email{manon.michel@cnrs.fr} \affiliation{Laboratoire de
  Math\'ematiques Blaise Pascal UMR 6620, CNRS, Université
  Clermont-Auvergne, Aubi\`ere, France.}

\begin{abstract}
  Event-Chain Monte Carlo (ECMC) methods generate continuous-time and
  non-reversible Markov processes which often display significant
  accelerations compared to reversible counterparts. However their
  generalization to any system may appear less straightforward. In
  this work, our aim is to distinctly define the essential symmetries
  that such ECMC algorithms must adhere to, differentiating between
  necessary and sufficient conditions. This exploration intends to
  delineate the balance between requirements that could be overly
  limiting in broad applications and those that are fundamentally
  essential. To do so, we build on the recent analytical description
  of such methods as generating Piecewise Deterministic Markov
  Processes (PDMP). Thus, starting with translational flows, we
  establish the necessary rotational invariance of the probability
  flows, along with determining the minimum event rate. This rate
  identifies with the corresponding infinitesimal Metropolis rejection
  rate. Obeying such conditions ensures the correct invariance for any
  ECMC scheme. Subsequently, we extend these findings to encompass
  schemes involving deterministic flows that are more general than
  mere translational ones. Specifically, we define two classes of
  interest of general flows: the \emph{ideal} and \emph{uniform-ideal}
  ones. They respectively suppresses or reduces the event rates. From
  there, we implement a comprehensive non-reversible sampling of a systems
  of hard dimers by introducing rotational flows,
  which are uniform-ideal. This implementation results in a speed-up
  of up to $\sim 3$ compared to the state-of-the-art ECMC/Metropolis
  hybrid scheme.
\end{abstract}

\maketitle

\section{Introduction}

The seminal Metropolis algorithm \cite{Metropolis_1953} was first
applied to hard-sphere systems, marking the genesis of Markov-chain
Monte Carlo (MCMC) methods. Over time, these methods have become an
ubiquitous tool in computational physics \cite{Frenkel_2001}. They
produce numerical evaluation of high-dimensional integrals by a
discrete summation over random configuration samples. Such samples are
realizations of a Markov process which explores the configuration
space of a system. In statistical physics, such high-dimensional
integrals often emerge from a stochastic description associated with a
Boltzmann probability distribution $\pi$. This distribution remains
invariant under the Markov process. The accuracy of such numerical
approximation is crucially linked on the ability of the Markov process
to produce a sequence of samples with minimal correlation
\cite{Janke_2002}. The Metropolis algorithm generates reversible
Markov processes, i.e. which obey the detailed balance
(e.g. $\pi_xP_{xy}=\pi_yP_{yx}$ for a discrete Markov chain with
transition probabilities $P$). The detailed balance is a stricter and sufficient
condition than the necessary global balance one (e.g. $\pi P=\pi$ for
a discrete chain) to ensure the invariance of the target
distribution. Such detailed-balance schemes most often rely on
rejections to target the correct invariant distribution.  Reversible
process then typically displays diffusive dynamics, most often leading
to excessively long correlation times \cite{Levin_2017}. This situation can
be further worsened in the presence of critical slowing down phenomena at
phase transitions \cite{Hohenberg_1977}, where the correlation length
diverges and spans the full system. Therefore, in order to alleviate
finite-size effects, a large community effort is devoted to the
developments of MCMC methods which exhibit better performance and
scalability with system sizes.

The efficient reversible clusters methods
\cite{Swendsen_1987,Wolff_1989} were formulated for lattice spin
systems thanks to thire spin-flip symmetry. In continuous particle
systems, such a natural involution is absent. This has prompted the
development of non-reversible MCMC methods known as Event-Chain Monte
Carlo (ECMC) \cite{Bernard_2009,Michel_2014}, which introduce an extra
variable $v$. ECMC in particle systems may be described as follows: a
particle follows a continuous-time ballistic trajectory set by $v$ up
to some \emph{event}; these events are ruled either by a jump process
reacting to the energy changes along the particle trajectory or
following some homogeneous rate independent from the physics (the
\textit{refreshment}); at an event, both the moving particle and the
parameter $v$ defining the ballistic trajectory are resampled by some
Markov kernel, initiating a ballistic move for a new particle. The
scheme then is rejection-free and relies on a control by the events of
the ballistic exploration to ensure the correct invariant
distribution. The observed accelerations have motivated the
applications of such methods to a large variety of systems, as
polymers \cite{Kampmann_2015}, continuous spins
\cite{Michel_2015,Nishikawa_2015} or all-atom simulations
\cite{Faulkner_2018}. Subsequent generalizations have primarily
focused on leveraging more comprehensive symmetries at the events, in
order to maintain the necessary control on the ballistic exploration,
while minimizing backtracking. Thus, starting with
the initial pairwise symmetry \cite{Michel_2014}, events can now
harness translational \cite{Harland_2017} and rotational
\cite{Michel_2020} symmetries or even the addition of some artificial
kinetic energy \cite{Klement_2019}.

While improving the control of the ballistic flow at the events has
been under much focus, the development of alternative ballistic flows
beyond standard translational updates has received smaller attention,
see for example \cite{Vanetti_2017,Bierkens_20}. Indeed, more involved
flow schemes may lead to an increased difficulty in computing the
event time in complex systems. However, the potential accelerations
could counterbalance this effect. More importantly, any ECMC
application to anisotropic particles, as dimers or more generally
molecules, necessitates the introduction of sequences of rotations to
achieve thermalization across all degrees of freedom. This is required
by the irreducibility condition, i.e. every admissible configurations
is reachable from any other admissible configurations, in order to get
a unique invariant measure. Because of the lack of rotational-flow
schemes, ECMC has so far been exclusively applied to tethered-type
\cite{Harland_2017,Hollmer_2022} or elastic
\cite{Harland_2017,Faulkner_2018} interactions in anisotropic
particles, so that translations can still thermalize the systems, or
has been coupled to some Metropolis scheme to generate rotational
moves \cite{Klement_2021}.  In addition to simulating all anisotropic
models, devising non-reversible general flows may lead to further
accelerations. It has been for instance observed that introducing
non-reversibility leads to accelerations that are greater as the
anisotropy of the particles is more pronounced \cite{Klement_2021}.
Finally, the design of generalized flows more generally raises the
question around the understanding of the necessary fundamental
conditions in ECMC versus the convenient sufficient choices, which is
reminiscent of the global balance vs detailed balance question.

In this work, we directly address this question, which allows us to
explicitly generalize the ECMC methods to more general ballistic
flows.  To do so, we rely on the characterization of ECMC as a
Piecewise Deterministic Markov process (PDMP)
\cite{Monemvassitis_2023} to precisely differentiate necessary from
sufficient conditions to ensure the invariance of the target
distribution. We do so first for the standard case of translational
flows. We establish the necessary rotational invariance of the
probability flows in the general case of local events and a
fundamental minimum event rate, which comes down to the corresponding
infinitesimal Metropolis rejection rate. This reveals that one of the
state-of-the-art ECMC schemes, known as the Forward variant or more
precisely the direct event kernel \cite{Michel_2020}, actually does
not require any further restrictive conditions to be implemented, as
was previously assumed.  We then consider the case of general flows,
where we generalize the necessary conditions of rotational invariance
and minimum event rate. Furthermore, we show that the Markov kernels
used in the translational case can be directly adapted to any
generalized flow. The minimum event rate still identifies with a
corresponding infinitesimal Metropolis rejection rate, which
encompasses as extra terms, beyond the physical infinitesimal energy
change, the divergence of the flows and the infinitesimal change of
the negative log probability of the variable $v$. Thus we define
\emph{ideal} flows, which exploit the possible compensation between
the different terms to reduce to 0 the minimum event rate for any
distribution. Moreover, we define \emph{uniform-ideal} flows, which
reduce the minimum event rate to its physical energy change component,
thus setting it to zero for uniform distributions. Our findings
underscore the potential of generalized flows to reduce event rates,
albeit possibly demanding more intricate flow integration. Indeed,
\emph{ideal} flows may be generally out of reach in terms of
algorithmic implementations, as they require some continuous
integration of the energy gradient. \emph{Uniform-ideal} flows can
however be more easily implemented, as they do not require such
integration and rely on deterministic evolutions independent from the
energy gradient. They include for instance the standard translational
but also the rotational flows.  We design and numerically benchmark
such rotational flows in both isotropic and anisotropic systems,
specifically bidimensional systems of hard spheres, fundamental MCMC
testbeds, and of hard dimers. In hard-sphere systems, it appears that
rotational flows retain good irreducibility property, regardless of
the refreshment rate. In hard-dimer systems, the symmetry of the
dimers itself interestingly imposes stringent conditions on the possible
deterministic flows. The introduction of irreversible rotations with
no backtracking moves leads, for the considered system size of 32
dimers, to an acceleration factor of up to 3 at the highest density
considered ($\rho=0.7$) compared to the state-of-the-art ECMC scheme
\cite{Klement_2021}.

We start by introducing in Section~\ref{sec:mcmc} the sampling of
particle systems by a standard Metropolis algorithm. We then study in
depth in Section~\ref{sec:ecmc} the ECMC method with standard
translational flow, and such directly in a PDMP framework which allows
to derive the necessary requirements as rotational invariance of the
probability flows and the minimum event rate. We also illustrate the
standard translational ECMC on isotropic particle systems.
Section~\ref{sec:flow} generalizes such study to more general
deterministic flows and we discuss in Section~\ref{sec:rot} the design
of rotational flows for hard spheres and for anisotropic bidimensional
dimers. Finally, we study the impact of the flow nature on the
numerical performances in a system of bidimensional hard spheres and
dimers in Section~\ref{sec:numerics}.

\section{Metropolis sampling for particle systems}
\label{sec:mcmc}
We consider a system composed of $N$ isotropic particles in a
$d$-dimensional periodic box of length $L$. A system configuration
$\bm{x}\in \mathcal{S}=(\mathbb{R}/(L\mathbb{Z}))^{dN}$ is entirely described by all the particle
positions $x_i$, i.e.  $\bm{x}=(\bm{x}_i)_{1\leq i \leq N}$.  We introduce
the notation $\bm{x}_{\setminus i}=((\bm{x}_j)_{j<i},(\bm{x}_j)_{j>i}$ for the
configuration of all particle positions but the $i$-th one. The
particles all interact in a repulsive pairwise manner following the
potential,
\begin{equation}
  U(\bm{x}) = \sum_{i<j} u(r(\bm{x}_i,\bm{x}_j)),
\end{equation}
with  $r$ the periodic distance on the torus. Furthermore, the system
may only admit a subset $\Omega\subset\mathcal{S}$ of valid
configurations, as is the case when dealing with particles with hard
cores of diameter $\sigma$ that cannot overlap.

Noting $\beta$ the inverse temperature, the system steady state then
follows the Boltzmann measure,
\begin{equation}
  \pi(\bm{x}) \propto \1_{\Omega}(\bm{x})\exp(-\beta U(\bm{x})),
  \end{equation}
  which we extend by continuity on $\partial \Omega$, and where $\1_{\Omega}(\bm{x})=1$ if $\bm{x}\in\Omega$ and 0 otherwise. In the
  following, we will set $\beta=1$ without loss of generality. A
  typical Metropolis algorithm sampling from $\pi$ consists in a
  proposal distribution $q$, verifying
  $\int_{\mathcal{S}}q(\bm{x}'|\bm{x})\d \bm{x}' = 1$ and commonly
  chosen to be a uniform increment of the position $\bm{x}_i$ of a random
  single sphere $i$ in $\bm{x}$, e.g.
  \begin{equation}
    q(\bm{x}'|\bm{x}) = \frac{1}{Nh^d}\sum_{i=1}^N \1_{\bm{x}_{\setminus i}}(\bm{x}'_{\setminus i})\1_{C_1}\left(\frac{\bm{x}_i-\bm{x}_i'}{h}\right)
    \end{equation}
    with $h>0$ some step amplitude and $C_1$ the centered unit
    hypercube of $\mathbbm{R}^d$ and, in an acceptance rate,
  \begin{equation}
    a(\bm{x}'|\bm{x})=\min\left(1,\frac{\pi(\bm{x}')}{\pi(\bm{x})}\right).
    \label{eq:Metroacc}
    \end{equation}
    The Markov chain generated by this Metropolis algorithm then
    follows the following kernel,
    \begin{multline}
        K(\bm{x},\bm{x}')=q(\bm{x}'|\bm{x})a(\bm{x}'|\bm{x}) \\\!+\! \Big(1\!-\!\int_{y \mathrlap{\in \mathcal{S}}}q(\bm{y}|\bm{x})a(\bm{y}|\bm{x})\d \bm{y}\Big)\1_{\{\bm{x}\}}(\bm{x}'),
    \label{eq:Metro}
  \end{multline}
  which leaves $\pi$ invariant, as it satisfies the detailed balance,
  \begin{multline}
    \pi(\d \bm{x}')K(\bm{x}',\d \bm{x})=\pi(\d \bm{x})K(\bm{x},\d \bm{x}').
    \label{eq:DB}
    \end{multline}
    It is however the global balance,
  \begin{equation}
    \int_{\bm{x}' \mathrlap{\in \mathcal{S}}} \pi(\d \bm{x}')K(\bm{x}',\d \bm{x})=\int_{\bm{x}' \mathrlap{\in \mathcal{S}}}\pi(\d \bm{x})K(\bm{x},\d \bm{x}'),
    \label{eq:GB}
  \end{equation}
  which is the necessary condition for the invariance of
  $\pi$. Various methods have then been devised to enhance efficiency
  by generating non-reversible processes which only obeys
  (\ref{eq:GB}), notably the Event-Chain Monte Carlo (ECMC) method
  \cite{Bernard_2009,Michel_2014}. However, the generalization of such
  methods to any systems, for instance to the sampling of anisotropic
  particles requiring rotations, is not straightforward. In contrast,
  the Metropolis algorithm can be adapted quite easily by considering
  the correct potential $U$ and by proposing steps that ensure
  irreducibility, as adding rotations for anisotropic
  particles. Aiming at devising such flow generalization for ECMC, we
  first introduce in the next section the standard ECMC method based
  on translations through their analytical characterization
  \cite{Monemvassitis_2023}. This approach allows for a precise
  differentiation between necessary symmetries and the imposed
  sufficient ones, akin to the distinction between global and
  detailed balances. From there, we discuss how to introduce
  generalized flows and especially explain how to generate rotations.

  \section{Event-Chain Monte Carlo with translational flow}
  \label{sec:ecmc}

\begin{figure*}
   	  \center \includegraphics{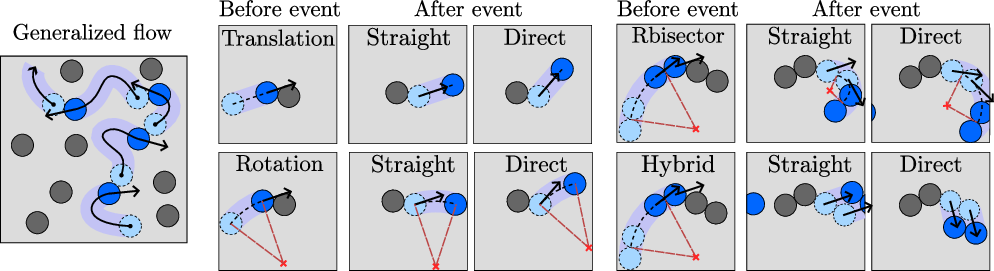}
          \caption{Representation of a generalized flow in an
            Event-Chain sampling of hard spheres ({\bf left}) and of
            different explicit uniform-ideal flows for spheres ({\bf
              middle}), namely translations and rotations, and for
            dimers ({\bf right}), namely bisector rotations and their
            hybrid variant combined with translations. Both the
            deterministic straight kernel and the stochastic direct
            kernel are illustrated.}
      \label{fig:schemes}
 	 \end{figure*}

  Contrary to the reversible and discrete-time Metropolis scheme,
  ECMC \cite{Michel_2014} generates a continuous-time and
  non-reversible Markov process. Considered as the infinitesimal limit
  of a discrete-time scheme, it then breaks detailed balance while
  still satisfying the global one. To do so, a lifting variable
  $\bm{v}\in\mathcal{V}$ following some chosen distribution $\mu$ is
  introduced so that the state space $\mathcal{S}$ is extended to
  $\mathcal{S}\times\mathcal{V}$. The extended state
  $(\bm{x},\bm{v})$ now follows the product measure $\pi \otimes
  \mu$. The lifting procedure \cite{CLP99,DHN00}, i.e. enlarging
    the state space to break reversibility while maintaining
    Markovianity, was introduced to speed up convergence of particular
    Markov chains (random walks on groups, ...). Here, the generated
  non-reversible process is then composed of ballistic updates of
  $\bm{x}$ set by $\bm{v}$, which is updated by a Markov kernel at
  domain boundaries (e.g. when two hard cores collides in a particle
  system) or at the events of a Poisson process whose rate depends on
  the infinitesimal potential $U$ increment. Thus, in particle
  systems, the generated process commonly comes down to updating a
  sphere position along some direction, until an event stemming from a
  pairwise interaction or hardcore collision with another sphere
  occurs. Then, this latter sphere's position is then the one being updated, in a
  billiard-like fashion as can be seen on
  Fig.~\ref{fig:schemes}. Then, on the extended state space
  $\mathcal{S}\times\mathcal{V}$, the former rejections in a
  Metropolis scheme are now transformed into an update of
  $\bm{v}$, making the scheme rejection-free.

  ECMC was initially built and justified through the infinitesimal
  limit of a discrete-time lifted Markov chain \cite{Michel_2014}. The
  process $(\bm{x}_t,\bm{v}_t)$ however identifies with a Piecewise
  Deterministic Markov process \cite{Davis_1984,Davis_1993}. This
  allows for a direct and efficient formalization directly at the
  continuous-space and -time level through its infinitesimal generator
  and boundary conditions. In particular, the invariance condition
  becomes straightforward. Following the PDMP characterization of ECMC
  detailed in \cite{Monemvassitis_2023}, we first express in terms of
  PDMP the ECMC process in a general setting, as represented in the
  algorithm \ref{algo:ECMC}. We then characterize the necessary
  symmetries required for the invariance of the target distribution
  and the sufficient symmetries more commonly applied to get explicit
  schemes. We finally illustrate it on the example of particle
  systems.

  \subsection{PDMP characterization}
  \label{sec:ecmc_pdmp}
  
         {\noindent \bf Generator and boundary kernel.} First, the
         generated process is characterized in the bulk, i.e. for
         $(\bm{x},\bm{v})\in\Omega\times\mathcal{V}$, by its generator
         $\mathcal{A}$, coding for the infinitesimal changes through
         the process on any observable $f$ (i.e.
         $\mathcal{A}f=\lim_{t\to
           0}\tfrac{1}{t}E_{\bm{x},\bm{v}}[f(\bm{x}(t),\bm{v}(t))-f(\bm{x},\bm{v})]$).
         In a general setting, the generator of the ECMC process may be written as

    \begin{equation}
      \mathcal{A}= \underbrace{\langle \bm{\phi}((\bm{x},\bm{v})) , \nabla \rangle}_{\text{Transport}} + \underbrace{\underbrace{\lambda(\bm{x},\bm{v})}_{\text{rate}}\big(\underbrace{Q((\bm{x},\bm{v}), \cdot)}_{\text{jump}}-\text{Id}\big)}_{\text{Events}},
    \label{eq:gen}
  \end{equation}
  where the deterministic flow $\bm{\phi}$ is such that
  $(\dot{\bm{x}}(t),\dot{\bm{v}}(t)) =
  \bm{\phi}(\bm{x}(t),\bm{v}(t))=\big(\bm{\phi}_x(\bm{x}(t),\bm{v}(t)),\bm{\phi}_v(\bm{x}(t),\bm{v}(t))\big)$
  during the ballistic phase. The rate $\lambda(\bm{x},\bm{v})$ rules
  the event times, at which the flow is controlled through the update
  of the lifting variable $\bm{v}$ by the Markov kernel $Q$ to another
  state $\bm{v}'\in\mathcal{V}$.  First, we consider the common case
  where the flow
  $\bm{\phi}(\bm{x},\bm{v})\!=\!\bm{\phi}^T(\bm{x},\bm{v})\!=\!(\bm{\phi}_x(\bm{v}),0)$
  identifies with a homogeneous translation of $\bm{x}$ uniquely set
  by a fixed $\bm{v}$. We discuss the impact of more general
  transports on the invariance conditions in the next section.

  In presence of hard validity constraints on $\Omega$, as hardcore
  interactions, the ECMC characterization is complete once added a
  boundary Markov kernel $Q_b$, which updates a state
  $(\bm{x},\bm{v})\in\partial\Omega\times \mathcal{V}$ that is exiting
  the bulk ($\langle n(\bm{x}),\bm{\phi}_x(\bm{v})\rangle > 0$,
  $n(\bm{x})$ being the local normal on the boundary, such states form
  up the set $\Gamma^+$) to an entering
  one ($\langle n(\bm{x}),\bm{\phi}_x(\bm{v}')\rangle < 0$, forming up the set $\Gamma^-$).\\

  {\bf \noindent Refreshment and irreducibility.} Irreducibility may
  require the introduction of a refreshment mechanism, which updates
  the lifting variable $\bm{v}$ according to its invariant
  distribution. This refreshment can be introduced at an exponential
  homogeneous rate, then appearing directly in the
  generator. Alternatively, it can be implemented through various
  schemes, including the commonly used fixed-time refreshment. The
  formalization of refreshment as a boundary effect
  \cite{Monemvassitis_2023} permits this flexibility. As the
  refreshment term self-cancels in the invariance computation, we omit
  its consideration in the subsequent discussion.
  \\
      
     {\bf \noindent Outputting samples.} As the generated process is
     continuous-time, observables of interest must be integrated along
     the full trajectory $(\bm{x}_t,\bm{v}_t)$ or averaged over an unbiased
     collection of samples $(\bm{x}_{\tau_n},\bm{v}_{\tau_n})_{\tau_n}$
     from the full trajectory. As the refreshment process, $\tau_n$
     can follow an exponential law or identify with some fixed
     value. In particular, if the vectors $\bm{\phi}_x(\bm{v})$ are of same norm,
     it is equivalent to output samples once a fixed length has been
     traveled over, which is the method used in most works. If
     $\bm{\phi}_x(\bm{v})$ can vary in norm, one should be careful to output
     samples at a fixed time or at a fixed length but renormalized by
     the norm of $\bm{\phi}_x(\bm{v})$.

\begin{widetext}
\begin{algorithm}
 \caption{ \label{algo:ECMC} Event-Chain Monte Carlo with general flows}
  \KwData{flow $\bm{\phi}$, event rate $\lambda$, Markov kernels $Q$, $Q_b$, refreshment time $T_{\text{Ref}} \geq 0$ and initial points $(\bm{x}_0,\bm{v}_0)$}
 Initialize $S_0 =0$ and a sequence of i.i.d~exponential random variables $(E_i)_{i \geq 1}$ with parameter $1$\\
\mbox{Set  $T_1^{\text{Ref}} = T_{\text{ref}}$}\hfill\mbox{\footnotesize\emph{Time before refreshment (can also follow some distribution \cite{Monemvassitis_2023})}}\\
 \For{$n \geq 0$}{  
   \mbox{Set $T_{n+1}^{\text{Ev}} = \inf\{ t \geq 0; \int_{0}^t \lambda\big((\bm{x}_{S_n},\bm{v}_{S_n}) + \int_0^s\bm{\phi}(\bm{x}_{S_n + u}, \bm{v}_{S_n + u})\d u\big) \d s \geq E_{n+1}\}$ \shortstack{\footnotesize\emph{Time before}\\\footnotesize\emph{event}}}\\
\mbox{Set $T_{n+1}^{\text{Bound}} = \inf\{ t \geq 0; \big((\bm{x}_{S_n},\bm{v}_{S_n}) + \int_0^t\bm{\phi}(\bm{x}_{S_n + u}, \bm{v}_{S_n + u})\d u\big)\not\in \Omega\}$ \shortstack{\footnotesize\emph{Time before boundary}}}\\
\mbox{Set $T_{n+1} = \min(T_{n+1}^{\text{Ev}},T_{n+1}^{\text{Bound}},T_{n+1}^{\text{Ref}})$}\hfill\mbox{\shortstack{\footnotesize\emph{Time before first resampling of $\bm{v}$}}}\\
\mbox{Set  $S_{n+1} = S_n + T_{n+1}$}\\
 \mbox{Set  $(\bm{x}_{S_{n+1}},\bm{v}_{S_{n+1}}) = (\bm{x}_{S_n},\bm{v}_{S_n}) + \int_0^{T_{n+1}}\bm{\phi}(\bm{x}_{S_n + u}, \bm{v}_{S_n + u})\d u$} \hfill\mbox{\shortstack{\footnotesize\emph{Update configuration and}\\\footnotesize\emph{lifting variable along $\bm{\phi}$}}}\\
 \uIf{$T_{n+1} = T_{n+1}^{\text{Ev}}$}{
   \mbox{ $\bm{v}_{S_{n+1}} \sim Q((\bm{x}_{S_{n+1}}, \bm{v}_{S_{n+1}}), \cdot)$} \hfill\mbox{\footnotesize\emph{Update the lifting variable according to the event kernel}}\\
         \mbox{Set   $T_{n+2}^{\text{Ref}} = T_{n+1}^{\text{Ref}} - T_{n+1}$}\hfill\mbox{\footnotesize\emph{Update the time before refreshment}} \\
       }
 \uElseIf{$T_{n+1} = T_{n+1}^{\text{Bound}}$}{
   \mbox{ $\bm{v}_{S_{n+1}} \sim Q_b((\bm{x}_{S_{n+1}}, \bm{v}_{S_{n+1}}), \cdot)$} \hfill\mbox{\footnotesize\emph{Update the lifting variable according to the boundary kernel}}\\
         \mbox{Set   $T_{n+2}^{\text{Ref}} = T_{n+1}^{\text{Ref}} - T_{n+1}$}\hfill\mbox{\footnotesize\emph{Update the time before refreshment}} \\
       }
   \uElse{
          \mbox{Set $v_{S_{n+1}} \sim \mu(\cdot)$}\hfill\mbox{\footnotesize\emph{Refresh the lifting variable}}\\
          \mbox{Set  $T_{n+2}^{\text{Ref}} = T^{\text{Ref}}$}\hfill\mbox{\footnotesize\emph{Update the time before refreshment}}
}
 }
\end{algorithm}
\end{widetext}     

  \subsection{Necessary and sufficient symmetries for
    invariance}
\label{sec:necessary_trans}
     
  {\noindent \bf Invariance.} The PDMP framework allows for a
        direct derivation of the invariance of the target $\pi$
        distribution,
\begin{equation}
  \int_{\Omega\times\mathcal{V}}\mathcal{A}f(\bm{x},\bm{v})\d\pi(\bm{x})\d\mu(\bm{v})=0,
  \label{eq:inv_gen}
\end{equation}
for test functions $f$ satisfying the boundary condition for
$(\bm{x},\bm{v'})\in\Gamma^-$,
\begin{equation}
  \int_{\mathcal{V}}\1_{\Gamma^+}((\bm{x},\bm{v}))Q_b((\bm{x},\bm{v}),
  \d\bm{v}')f(\bm{x},\bm{v})\d\mu(\bm{v})=f(\bm{x},\bm{v}').
  \label{eq:bound_cond}
  \end{equation}
  This necessary symmetry on the test functions $f$ can be
    rewritten as the condition $Q_bf-f=0$, which can be understood as
    the properly defined condition imposed by a jump mechanism with an
    \emph{infinite} rate. If it was not a boundary term and had a
    finite rate $\lambda'$, it would be expressed in the generator as
    $\lambda' (Q_bf-f)$. This is for instance the difference between
    events in hard-sphere (\emph{infinite} rate and described by a
    boundary kernel) and soft-sphere systems (finite rate event), for
    a longer introduction see \cite{Monemvassitis_2023}. An
  integration by parts on the transport term shows the necessary
  compensation in the bulk between the transport and the
  event terms, giving in the fixed translation case,
\begin{multline}
    \big(\langle \bm{\phi}_x(\bm{v}), -\nabla U(\bm{x})\rangle + \lambda(\bm{x},\bm{v})\big)\mu(\bm{v})
 \\   = \int_{\mathrlap{\mathcal{V}}}\lambda(\bm{x},\bm{v}')Q((\bm{x},\bm{v}'), \bm{v})\d\mu(\bm{v}'),
  \label{eq:inv_gen_dev}
  \end{multline}
  and, using the boundary condition (\ref{eq:bound_cond}), the
  transport cancellation by the boundary kernel at the boundary, for
  $(\bm{x},\bm{v})\in\partial\Omega\times\mathcal{V}$,
  \begin{multline}
    \langle \bm{\phi}_x(\bm{v}), n(\bm{x})\rangle_- \mu(\bm{v}) 
 \\   \!=\! \int_{\mathrlap{\mathcal{V}}}Q_b((\bm{x},\bm{v}'), \bm{v})\langle \bm{\phi}_x(\bm{v}'), n(\bm{x})\rangle_+\d\mu(\bm{v}'),
  \label{eq:inv_gen_dev-b}
  \end{multline}
  where the underscript $x_+$ denotes the positive part of $x$
    (and $x_-=(-x)_+$). The transport cancellation at the boundary can
    again be understood as the \emph{infinite} rate limit of the
    transport cancellation by the event jump mechanism in the bulk.

    Similar to the global balance for discrete-time Markov chains,
    both conditions (\ref{eq:inv_gen_dev}) and
    (\ref{eq:inv_gen_dev-b}) are the only necessary and sufficient
    ones to ensure the correct invariance. Similarly to the first
    discrete-time MCMC schemes which were detailed-balanced and obeyed
    stricter sufficient conditions, the current MCMC schemes based on
    PDMP actually satisfy more restrictive conditions than the one
    necessary for invariance. We now more precisely derive the
    necessary conditions stemming from (\ref{eq:inv_gen_dev}) and
    (\ref{eq:inv_gen_dev-b}).
    \\

  \noindent {\bf Necessary symmetries.}  First, a direct necessary
  condition on the choice of $\mu$ and $\bm{\phi}$ is the one of the
  conservation of the probability flows, obtained by an integration
  over $\bm{v}$  in (\ref{eq:inv_gen_dev}) or
  (\ref{eq:inv_gen_dev-b}),
\begin{equation}
  \def\arraystretch{1.4}
\left\{  \begin{array}{l}
  \int_{\mathcal{V}}  \langle \bm{\phi}_x(\bm{v}), \nabla U(\bm{x})\rangle \d\mu(\bm{v}) 
    = 0\\
  \int_{\mathcal{V}}  \langle \bm{\phi}_x(\bm{v}), n(\bm{x})\rangle \d\mu(\bm{v}) 
    = 0\end{array}\right.,
  \label{eq:conservation}
\end{equation}
i.e.,
\begin{equation}
  \def\arraystretch{1.4}
\left\{  \begin{array}{l}
  \int_{\mathcal{V}}  \langle \bm{\phi}_x(\bm{v}), \nabla U(\bm{x})\rangle_+ \d\mu(\bm{v})
    \!=\! \int_{\mathcal{V}}  \langle \bm{\phi}_x(\bm{v}), \nabla U(\bm{x})\rangle_- \d\mu(\bm{v}) \\
  \int_{\mathcal{V}}  \langle \bm{\phi}_x(\bm{v}), n(\bm{x})\rangle_+ \d\mu(\bm{v}) 
    \!=\!   \int_{\mathcal{V}}  \langle \bm{\phi}_x(\bm{v}), n(\bm{x})\rangle_- \d\mu(\bm{v}) \end{array}\right.,
  \label{eq:mu_balance}
\end{equation}
Thus, the distribution and the flow set by the lifting variable
$\bm{v}$ must ensure that there is a balance between probability flows
increasing or decreasing along $\nabla U$, i.e. the gradient of
potential $U$, or along $n(\bm{x})$ on the boundary.  This necessary
flow symmetry (\ref{eq:mu_balance}) underlines that, when the time
reversibility is broken, it is replaced by another key symmetry
ensured by $\bm{\phi}$ and $\mu$ and whose origin is deeply rooted in the
conservation of probability flow itself.

Furthermore, such symmetry plays a central role in the transport-event
compensation. First, as the right-hand term of (\ref{eq:inv_gen_dev})
is positive, the rate $\lambda$ must obey the following condition for
all $(\bm{x},\bm{v})\in\Omega\times\mathcal{V}$,
  \begin{equation}
    \lambda(\bm{x},\bm{v}) - \langle \bm{\phi}_x(\bm{v}), \nabla U(\bm{x})\rangle \geq 0,
    \label{eq:min_rate}
    \end{equation}
so that the choice realizing the smallest event rate possible is,
\begin{equation}
  \lambda_M(\bm{x},\bm{v}) = \langle \bm{\phi}_x(\bm{v}), \nabla U(\bm{x}) \rangle_+.
  \end{equation}
  It corresponds to the rejection rate of the equivalent infinitesimal
  lifted Metropolis scheme. When constructing ECMC as such
  infinitesimal limit, this choice seems natural \cite{Michel_2014}
  but (\ref{eq:min_rate}) further shows that it is indeed the
  minimal possible rate.

  Considering now the conditions on the event Markov kernel $Q$, we
  obtain by setting $\lambda$ to $\lambda_M$ in (\ref{eq:inv_gen_dev}), 
\begin{multline}
   \langle \bm{\phi}_x(\bm{v}), \nabla U(\bm{x})\rangle_-\mu(\bm{v})  \\ \!=\! 
          \int_{\mathcal{V}}\d \bm{v}' \mu(\bm{v}')\langle \bm{\phi}_x(\bm{v}'), \nabla U(\bm{x})\rangle_+Q((\bm{x},\bm{v}'), \bm{v}).
  \label{eq:balance_Q}
\end{multline}
Thus, the kernel $Q$ can be set to any kernel which leave invariant,
up to a \emph{flip}, the distribution
$\langle \bm{\phi}_x(\bm{v}), \nabla
U(\bm{x})\rangle_+\mu(\d\bm{v})$. A valid explicit choice then is the
direct sampling,
\begin{equation}
  Q_{\text{Dir}}((\bm{x},\bm{v}'), \d\bm{v}) = \frac{\langle \bm{\phi}_x(\bm{v}), \nabla U(\bm{x})\rangle_-\mu(\d\bm{v})}{\int  \langle \bm{\phi}_x(\bm{v}), \nabla U(\bm{x})\rangle_+\mu(\d\bm{v})},
  \label{eq:direct_pick}
  \end{equation}
  which correctly sums up to $1$ thanks to the necessary flow symmetry
  (\ref{eq:mu_balance}) and does not depend on the norm of the local
  gradient $\nabla U(\bm{x})$. Such kernel can be understood as a
  direct pick from the \emph{event} distribution
  ($\langle \bm{\phi}_x(\bm{v}), \nabla
  U(\bm{x})\rangle_+\mu(\d\bm{v})$) combined with some
  \emph{flip}.  Such derivation is similar regarding the boundary
  condition (\ref{eq:inv_gen_dev-b}) and $Q^b$ can be chosen as $Q$
  once updated the reference vector from $\nabla U(\bm{x})$
  to $n(\bm{x},\bm{v})$.

  This type of kernel was already introduced in the Forward
  Event-Chain generalization \cite{Michel_2020} by explicitly assuming
  that $\bm{\phi}_x(\bm{v})\mu(\bm{v})$ is rotationally-invariant in
  order to derive (\ref{eq:mu_balance}). We showed that the symmetry
  of probability flows around the gradient (\ref{eq:mu_balance})
  actually is as necessary as the conservation condition
  itself. Therefore, this demonstrate that such direct sampling is
  always possible without further assumption. However, the advantage
  of imposing a rotationally-invariant property lies in an easier and
  general implementation of the kernel (\ref{eq:direct_pick}). Indeed,
  the dependence on $\bm{x}$ then only impacts the decomposition of
  $\bm{v}$ into a parallel component along $\nabla U(\bm{x})$, and an
  orthogonal one, whose values can be resampled independently from
  $\bm{x}$. More generally, restricting $\mu$ to some choice allowing
  to easily define the \emph{flip} mentioned earlier allows some
  explicit construction we now address.
  \\
  
  \noindent {\bf Sufficient symmetries.} The need to derive explicit
  and general choices for $\lambda$, $Q$ and $Q_b$ has led to the
  development of schemes satisfying stricter sufficient conditions but
  enabling easier numerical implementation. Similarly, the use of
  detailed balance has historically facilitated the implementation of
  discrete-time Markov chains. Actually, the choice of imposing a
  product measure $\pi\otimes\mu$ is already a stricter choice than
  the necessary condition that $\pi$ is the marginal of the extended
  measure, but it definitely simplifies the task of designing a
  correct scheme. We now discuss the most commonly used schemes.

First, we more generally decompose $\lambda$ as,
  \[ \lambda(\bm{x},\bm{v}) = \alpha(\bm{x},\bm{v}) + \langle
    \bm{\phi}_x(\bm{v}), \nabla U(\bm{x}) \rangle_+,
  \]
  with $\alpha(\bm{x},\bm{v}) \geq 0$ some excess rate. The condition
  (\ref{eq:inv_gen_dev}) is now decomposed as,
\begin{equation}
  \begin{aligned}
   \langle \bm{\phi}_x(\bm{v}), \nabla U(\bm{x})\rangle_-\mu(\bm{v})   \!=\! 
           \\
          \int_{\mathcal{V}}\d \bm{v}' \mu(\bm{v}')\langle \bm{\phi}_x(\bm{v}'), \nabla U(\bm{x})\rangle_+Q_\phi((\bm{x},\bm{v}'), \bm{v}),
    \end{aligned}           
  \label{eq:balance_phi}
\end{equation}
and,
\begin{equation}
     \alpha(\bm{x},\bm{v})\mu(\bm{v})\!=\!
     \int_{\mathcal{V}}\d \bm{v}
     ' \mu(\bm{v}')\alpha(\bm{x},\bm{v}')Q_\alpha((\bm{x},\bm{v}'), \bm{v})
  \label{eq:balance_a}
\end{equation}
with the Markov kernels $Q_\phi$ and $Q_\alpha$ so that,
\begin{multline}
    Q((\bm{x},\bm{v}'), \bm{v}) \!=\! 
  \frac{\langle \bm{\phi}_x(\bm{v'}), \nabla U(\bm{x})\rangle_+}{\lambda(\bm{x},\bm{v}')}Q_\phi((\bm{x},\bm{v}'), \bm{v})
\\  +\frac{\alpha(\bm{x},\bm{v}')}{
    \lambda(\bm{x},\bm{v}')}Q_\alpha((\bm{x},\bm{v}'), \bm{v}). 
\label{eq:Q_structure}
\end{multline}
The kernel $Q_\phi$ can be set to any kernel obeying
(\ref{eq:balance_Q}), as $Q_{\text{Dir}}$ (\ref{eq:direct_pick}).  The
Markov kernel $Q_\alpha$, stemming from the unnecessary excess rate
$\alpha$, can be any kernel which leaves
$\alpha(\bm{x},\bm{v})\mu(\bm{v})$ invariant, as a direct or
Metropolis-like sampling following
$\alpha(\bm{x},\bm{v})\mu(\bm{v})$. Common choices for $\alpha$ are
motivated by simplifying the computation of event times. For instance,
it can be done by the thinning of the actual minimal rate,
$Q_\alpha((\bm{x},\bm{v}),\d\bm{v}')$ then identifying with
$\delta(\bm{v}-\bm{v}')\d\bm{v}'$ (no real event is actually
sampled). Another method involves the factorization of the
minimal rate along the components of the gradient $\nabla U$,
such as along interaction terms, which may allow for some inversion
sampling of the event times.

Interestingly, in the case of factorized rates, writing,
\begin{multline}
    \alpha(\bm{x},\bm{v}) \!=\! \sum_i\langle  f_i(\bm{x},\bm{v})\rangle_+\!-\! \big\langle \sum_i f_i(\bm{x},\bm{v})\big\rangle_+
\\    \!=\! \sum_i\langle  f_i(\bm{x},\bm{v})\rangle_-\!-\! \big\langle \sum_i 
    f_i(\bm{x},\bm{v})\big\rangle_-,\notag
    \end{multline}
with $\{f_i\}$ the factors so that
$$\sum_i f_i(\bm{x},\bm{v})\!=\!\langle \bm{\phi}_x(\bm{v'}), \nabla U(\bm{x})\rangle,$$ we obtain by setting $Q_\phi$ and $Q_\alpha$
to the same kernel $Q_{\text{Fact}}$, as usually done, the condition,
\begin{multline}
   \sum_i\langle  f_i(\bm{x},\bm{v})\rangle_-\mu(\bm{v})  \\ \!=\! 
           \int_{\mathcal{V}}\d \bm{v}' \mu(\bm{v}') \sum_i\langle  f_i(\bm{x},\bm{v}')\rangle_+
           Q_{\text{Fact}}((\bm{x},\bm{v}'), \bm{v}),
  \label{eq:balance_fact_Q}
\end{multline}
which necessarily requires the factorized symmetry for flow
conservation,
\begin{equation}
  \begin{aligned}
        \sum_i \int_{\mathcal{V}}\mu(\d \bm{v)}  \langle  f_i(\bm{x},\bm{v})\rangle_-\!=\! 
          \sum_i  \int_{\mathcal{V}}\mu(\d \bm{v}) \langle  f_i(\bm{x},\bm{v})\rangle_+,
    \end{aligned}           
  \label{eq:balance_fact_glob}
\end{equation}
which comes down to (\ref{eq:mu_balance}), so that it is always
possible to set $Q_{\text{Fact}}$ to the direct variant
\begin{equation}
  Q^{\text{Glob}}_{\text{Fact,Dir}}((\bm{x},\bm{v}'), \d\bm{v}) = \frac{\sum_i\langle f_i(\bm{x},\bm{v})\rangle_-\mu(\d\bm{v})}{\int  \sum_i\langle f_i(\bm{x},\bm{v})\rangle_+\mu(\d\bm{v})}.
  \label{eq:direct_pick_fact_glob}
\end{equation}
However, it may prove interesting to consider $\bm{\phi}$ and $\mu$ so
that the condition (\ref{eq:balance_fact_glob}) is achieved in a more
restrictive manner thanks to some detailed symmetry, as,
\begin{equation}
  \begin{aligned}
\forall i,       \int_{\mathcal{V}}\mu(\d \bm{v)}  \langle  f_i(\bm{x},\bm{v})\rangle_-\!=\! 
            \int_{\mathcal{V}}\mu(\d \bm{v}) \langle  f_i(\bm{x},\bm{v})\rangle_+.
    \end{aligned}           
  \label{eq:balance_fact}
\end{equation}
This symmetry allows to treat every factor independently and to
consider the following kernel,
\begin{equation}
  \begin{aligned}
    Q^{\text{Det}}_{\text{Fact,Dir}}((\bm{x},\bm{v}'), \d\bm{v}) \!=\! \frac{\sum_i\langle f_i(\bm{x},\bm{v})\rangle_+\mu(\d\bm{v})Q_i}{\sum_i\int  \langle f_i(\bm{x},\bm{v})\rangle_+\mu(\d\bm{v})}\\
\text{with,}\    Q_i((\bm{x},\bm{v}'), \d\bm{v}) = \frac{\langle f_i(\bm{x},\bm{v})\rangle_-\mu(\d\bm{v})}{\int  \langle f_i(\bm{x},\bm{v})\rangle_+\mu(\d\bm{v})}.
    \end{aligned}
  \label{eq:direct_pick_fact}
\end{equation}
Drawing on
the lines of more restrictive and detailed symmetry for the flow conservation as in
(\ref{eq:balance_fact}) and of a particular role played by some
\emph{flip}, most explicit schemes, apart the direct kernel, are
building on some \emph{flip} mapping $F_x:\mathcal{V}\to\mathcal{V}$
present in flow $\phi$ and $\mu$, as,
\begin{equation}
  \def\arraystretch{1.4}
  \left\{
    \begin{array}{l}
      \langle \bm{\phi}_x(\bm{v}), \nabla U(\bm{x})\rangle \!=\! \!-\!\langle
      \bm{\phi}_x(F_x(\bm{v})), \nabla U(\bm{x})\rangle\\
      \mu(\bm{v})=\mu(F^{-1}_x(\bm{v}))
      \end{array}
    \right..
  \label{eq:flip}
\end{equation}
Such symmetry is sufficient to meet the flow conservation condition
(\ref{eq:conservation}) in a detailed manner. To get such mapping,
$\mu$ is typically chosen uniform or Gaussian so that
$\bm{\phi}_x(\bm{v})$ is rotationally invariant. Then, $F_x$ typically
codes at the level of $\bm{\phi}_x(\bm{v})$ for a full flip, a
reflection across the potential gradient $\nabla U$ or some
other particular symmetry exploiting the ones of $\nabla U$ as the
pairwise mirror symmetry or translational invariance in particle
systems. In particular, it is the existence and exploitation of such
underlying symmetry which has allowed to design deterministic Markov
kernels of the type $\delta(F_x(\bm{v})-\bm{v}')\d\bm{v}'$, in
comparison to the direct kernels previously discussed.
\\

  \noindent{\bf Further generalizations.} The necessary flow
  symmetry (\ref{eq:conservation}) can actually be alleviated when
  considering more general processes. From a general
  perspective, it is most of the times not possible to obtain without
  rejections direct samples from $\pi$, except when leveraging known
  symmetries of $\pi$ as is done in overrelaxation moves in
  spin systems. Therefore, focus has been on developing Markov kernel
  $Q$ that solely updates the lifting variable $\bm{v}$. We here consider
  more comprehensive kernels that potentially updates both the
  physical $\bm{x}$ and lifting $\bm{v}$ variables.

  First, the invariance condition then yields, in the bulk,
\begin{multline}
    \big(\langle \bm{\phi}_x(\bm{v}), \nabla \ln\pi(\bm{x})\rangle + \lambda(\bm{x},\bm{v})\big)\mu(\bm{v}) \pi(\bm{x})
\\    = \int_{\mathrlap{\Omega\times\mathcal{V}}}\lambda(\bm{x'},\bm{v}')Q((\bm{x'},\bm{v}'), (\bm{x},\bm{v}))\d\pi(\bm{x'})\d\mu(\bm{v}'),
  \label{eq:inv_gen_dev_x}
  \end{multline}
    and at the boundary,
  \begin{equation}
  \begin{aligned}
    \langle \bm{\phi}_x(\bm{v}), n(\bm{x})\rangle_- \mu(\bm{v}) \pi(\bm{x})\ \ \ \ \ \ \ \  \ \ \ \ \ \ \  \  \  \ \ \  \ \ \ \ \ \ \ \  \ \ \ \ \ \ \ \ \ \ \\
    \!=\! \int_{\mathrlap{\partial\Omega\times\mathcal{V}}}Q_b((\bm{x'},\bm{v}'), (\bm{x},\bm{v}))\langle \bm{\phi}_x(\bm{v}'), n(\bm{x'})\rangle_+\d\pi(\bm{x'})\d\mu(\bm{v}').
    \end{aligned}
  \label{eq:inv_gen_dev-b_x}
  \end{equation}
  So that the conservation of the probability flows actually require,
  by integrating over $(\bm{x},\bm{v})$ either in
  (\ref{eq:inv_gen_dev_x}) or (\ref{eq:inv_gen_dev-b_x}),
\begin{equation}
  \int_{\partial\Omega\mathrlap{\times\mathcal{V}}}  \langle \bm{\phi}_x(\bm{v}), n(\bm{x})\rangle \d\pi(\bm{x})\d\mu(\bm{v})=0. 
  \label{eq:conservation-x}
\end{equation}
This condition is a global one, compared to the local
(\ref{eq:conservation}), and is non-restrictive in the case where
$\partial\Omega=\emptyset$ or when $\pi$ already admits some symmetry
leading to, for any $\bm{e}\in\mathcal{S}$,
\[\int_{\partial\Omega}   \langle\bm{e}, n(\bm{x})\rangle \d\pi(\bm{x})=0,\]
as is the case of the particle systems presented in the previous
section, which presents the pairwise mirror symmetry
($\nabla_{x_i}
r(\bm{x}_i,\bm{x}_j)=-\nabla_{x_j}r(\bm{x}_i,\bm{x}_j)$). Otherwise,
the conservation condition imposes on the choice of the flow
$\bm{\phi}$ and distribution $\mu$ a global symmetry of the exit
($\langle\pi(\bm{x})\mu(\bm{v})\phi_x(\bm{v}),n(\bm{x})\rangle_+$) and
entering probability flow
($\langle\pi(\bm{x})\mu(\bm{v})\phi_x(\bm{v}),n(\bm{x})\rangle_-$)
along the boundary. Compared to reversible schemes, this condition
also directly stems from the non-reversible continuous-time ballistic
component of the process, which pushes it up to the boundary
$\partial\Omega$.

Eventually, the conservation or equivalently symmetry condition
(\ref{eq:conservation-x}) could furthermore be suppressed if the
Markov kernel $Q$ (resp. $Q_b$) was even more general and for instance
allowed to propose jumps from the bulk (resp. the boundary) to the
boundary (resp. the bulk).

\subsection{Translational ECMC for isotropic particle systems}
\label{sec:ecmc_part}
We now present commonly-used ECMC schemes in particle systems, as illustrated in Fig.~\ref{fig:schemes}. They are characterized by:\\
    
\noindent {\bf Transport.}  The lifting variable $\bm{v}$ identifies
with a tuple $(\bm{e},i)$ of a vector $\bm{e}$ in $\mathbb{R}^d$ and
the label $i\in\setind$ of the updated sphere along said
vector. Standard choices for $\mu$ is  a product measure
$\mu_e\otimes\mu_n$, with $\mu_e$ the uniform distribution over
$[\bm{u}_1,\mydots,\bm{u}_d]$ (moves along the canonical basis of
$\mathbb{R}^d$, as done in \cite{Bernard_2009,Michel_2014}) or over
$\mathbb{S}^{d-1}$ ($\bm{e}$ is some unit vector, as done in
\cite{Harland_2017,Michel_2020}) or a Normal distribution over
$\mathbb{R}^d$ (as used in \cite{Klement_2019}) and $\mu_n$ the
uniform distribution over $\setind$.
    
   The deterministic flow $\bm{\phi}$ identifies with the translational flow
  $\bm{\phi}^T$ which updates $(\bm{x},\bm{v})$ by a translation of the
  $i$-th sphere along $\bm{e}$, i.e. 
  \begin{equation}
    \bm{\phi}(\bm{x},(\bm{e},i)) \!=\! ((0,\mydots,0, \underbrace{\bm{e}}_{i-\text{th}},0,\mydots,0), (0,0)).
    \label{eq:flow_T}
  \end{equation}
  The conservation condition (\ref{eq:conservation}) is met, in
  particular thanks to the pairwise symmetry in the case of $\mu_e$
  the uniform distribution over the canonical basis.
  \\
  
  \noindent {\bf Event rate.} The rates are factorized along the pairwise interactions, i.e.,
      \begin{equation}
      \lambda(\bm{x},\bm{v})Q((\bm{x},\bm{v}), \cdot)\!=\! \sum_{j\neq i}\lambda_{ij}(\bm{x},\bm{v})Q_{j}((\bm{x},\bm{v}), \cdot),
      \label{eq:rate}
    \end{equation}
    with,
  \begin{multline}
      \lambda_{ij}(\bm{x},(\bm{e},i)) =\tfrac{1}{2}\langle  \nabla u(r(\bm{x}_i,\bm{x}_j)), \bm{e} \rangle_+
\\      =\tfrac{1}{2}| u'(r(\bm{x}_i,\bm{x}_j))|\langle \bm{n}_{ji}, \bm{e} \rangle_+,
    \label{eq:rates}
  \end{multline}
  with the pairwise normalized gradient
  $\bm{n}_{ij}(\bm{x})=2\nabla_{\bm{x}_i}r(\bm{x}_i,\bm{x}_j)=-\bm{n}_{ji}(\bm{x})$. Such
  superposition allows to treat every interaction independently. It
  can also ease the sampling of the Poisson process, either by
  inversion sampling \cite{Michel_2014} or by thinning
  \cite{Kapfer_2015}, as mentioned previously. The factorized flow
  conservation condition (\ref{eq:balance_fact}) is also met, making
  it possible to consider a kernel $Q$ similar to
  (\ref{eq:direct_pick_fact}).
  \\
  
  \noindent {\bf Markov kernel.} Thanks to the existence of a mapping
  $F_{\bm{x}}$ as in (\ref{eq:flip}) consisting in exchanging particle
  labels, it is possible to design deterministic kernels for
  $(Q_j)_{1\leq j\leq N}$, as the straight $Q^S$ or reflection kernel
  $Q^R$ \cite{Bernard_2009,Michel_2014},
    \begin{equation}
	\label{eq:straight_kernel}
\def\arraystretch{1.4}
      \left\{      \begin{array}{l}
                     Q^S_j((\bm{x},(\bm{e},i)),(\bm{e}',i'))=\1_{\{j\}}(i')\1_{\{\bm{e}\}}(\bm{e}')\\
                     Q^R_j((\bm{x},(\bm{e},i)),(\bm{e}',i'))=\1_{\{j\}}(i')\1_{\{R_x(\bm{e})\}}(\bm{e}'),\\                     
               \end{array}\right.
      \end{equation}
      with
      $R_x(\bm{e}) = -\bm{e}+2\langle \bm{n}_{ij}(\bm{x}),
      \bm{e}\rangle \bm{n}_{ij}(\bm{x})$. At an event involving the moving
      sphere $i$ and fixed one $j$, such kernels update the moving
      sphere from $i$ to $j$, either by keeping the same direction $\bm{e}$
      or reflecting it.

      Now, in the case of a rotationally-invariant $\mu_e(\cdot)$, as
      the uniform distribution over $\mathbb{S}^{d-1}$ or some Normal
      distribution over $\mathbb{R}^d$, the direct kernel variant
      (\ref{eq:direct_pick_fact}) is actually explicit
      \cite{Michel_2020}. By expressing $\bm{e}$ in spherical
      coordinates with $\bm{n}_{ij}(\bm{x})$ as a polar axis
      ($\langle \bm{n}_{ij}(\bm{x}), \bm{e} \rangle = r\cos\theta_1$),
       \[\mu_e(\d \bm{e}) \propto r^{d-1}\mu_{e,r}(\d r)\sin^{d-2}(\theta_1)\mydots\sin(\theta_{d-2})\d \theta_1\mydots\d\theta_{d-1},\]
       it yields the following event distribution,
       \begin{multline}
         \notag
        \int\1_A(\bm{e})\langle \bm{e}, \bm{n}_{ji}\rangle_+ \mu_e(\d \bm{e}) \\\propto \int_0^1 b^{d-2}\d b\int_{\mathbb{R}^+} r^{d-1}\mu_{e,r}(\d r) \1_{A}((r\sqrt{1-b^2}\bm{n}_{ij}(\bm{x}) \\\!+\! rb\tfrac{\bm{e}-\langle \bm{n}_{ij}(\bm{x}),\bm{e}\rangle \bm{n}_{ji}(\bm{x})}{||\bm{e}-\langle \bm{n}_{ij}(\bm{x}),\bm{e}\rangle||},i)).
          \end{multline}
        Thus, in the case of the uniform distribution over the
        hypersphere, a pair $(\bm{e},i)$ where $\bm{e}$ is uniformly
        oriented in the hyperplane
        $\text{Span}\{\bm{n}_{ij}(\bm{x})\}^\perp$ and admits $\arcsin (b)$
        as a polar angle to $\bm{n}_{ji}(\bm{x})$, follows the event
        distribution. In the case of a Gaussian distribution,
        $\langle \bm{e}, \bm{n}_{ji}(\bm{x}) \rangle$ should follow a
        $\chi2$ law and the other components of $\bm{e}$ the usual
        Gaussian one.

        In the following, we will consider the uniform distribution
        over $\mathcal{S}^{d-1}$ and the corresponding following
        direct kernel, sampling from the event distribution combined
        with the label flip,
      \begin{multline}
		\label{eq:direct_kernel}
        Q_j^{\text{Dir}}((\bm{x},(\bm{e},i)),A)\\ =\int_0^1 b^{d-2}\d b \1_{A}\Big(\big(\sqrt{1-b^2}\bm{n}_{ji}(\bm{x}) + b\tfrac{\bm{e}-\langle \bm{n}_{ij}(\bm{x}),\bm{e}\rangle \bm{n}_{ij}(\bm{x})}{||\bm{e}-\langle \bm{n}_{ij}(\bm{x}),\bm{e}\rangle||},j\big)\Big),
        \end{multline}
        which transfers the ballistic move from the sphere $i$ to the
        $j$-th one, keeps the same direction in the hyperplane orthogonal to
        $\bm{n}_{ij}(\bm{x})$ but, compared to the deterministic kernels, update
        the polar angle directly from its steady-state distribution,
        via a sampling of $b\in[0,1]$, typically by
        inversion sampling, as,
        \[ b = \nu^{1/(d-1)},\ \text{with}\ \nu \sim
          \text{Unif}(0,1).\]

   Finally, out of completeness, in the case of the
        uniform distribution over the canonical basis, a direct
        sampling is enforced by,
        \begin{multline}
            Q_j^{\text{Can}}((\bm{x},(\bm{e},i)),\d \bm{e}')\\\!=\!\1_{\{u_k\}_k}(\bm{e}')
           \frac{\langle \bm{n}_{ji}(\bm{x}),\bm{e}'\rangle_+}{\sum_{k=1}^d|\langle \bm{n}_{ij}(\bm{x}),\bm{u}_k\rangle|} \d \bm{e}',
        \end{multline}
        which can be sampled by inversion sampling.\\
        
        \noindent{\bf Boundary kernel.} A boundary kernel $Q_b$ is
        necessary to take into account the hardcore interactions. As
        detailed in \cite{Monemvassitis_2023} for particle system and
        more generally in the previous subsection, the boundary kernel
        can be chosen to identify with $Q$. Simply put, hardcore
        interactions can be understood as soft ones but with diverging
        $\lambda_{ij}\to\infty$.\\

\section{Generalized flow in ECMC}
\label{sec:flow}

\subsection{Necessary and sufficient symmetries for invariance}

Since its first developments, the translational flow $\bm{\phi}^T$ has
always been the one implemented in ECMC sampling. However, the PDMP
formalism allows for more general flow: if the ODE
$\frac{d(\bm{x}_t,\bm{v}_t)}{dt}=\bm{\phi}(\bm{x}_t,\bm{v}_t)$ with
differentiable drift $\bm{\phi}$ defines a c\`adl\`ag (in time)
$(\bm{\phi}_t)_{t\ge0}$ deterministic flow satisfying the semigroup
property ($\bm{\phi}_{t+s}=\bm{\phi}_t\circ\bm{\phi}_s$), then the
drift part of the generator (\ref{eq:gen}) is well defined. It may
however impact the conservation condition and the balance required for
invariance between the transport and event/boundary
contributions.\\

{\noindent \bf Invariance.} Indeed, in the general case, the
integration by part on the transport formally generates additional
terms as the divergence term $\nabla\cdot \bm{\phi}$ and derivative
$\langle \bm{\phi}, \nabla \mu(\bm{v})\rangle$, see for example
\cite{Vanetti_2017} and as we derive and extend the invariance
condition to constrained domain as imposed by hardcore interactions
below.  In the following, with a slight abuse of notation when
denoting $\nabla_v f(\bm{x},\bm{v})$, the gradient operator concerns
only the continuous part of $\bm{v}$, which is the only part of the
lifted variable that could be subject to the flow
$\bm{\phi}(\bm{x},\bm{v})$, However the flow $\bm{\phi}$ may depend
both on the continuous and discrete parts of the lifting variable
$\bm{v}$, e.g. when $\bm{v}$ denotes the velocity and label of the
active particle. By integration by parts,
\begin{multline}
  \int_{\Omega\mathrlap{\times\mathcal{V}}}\langle \bm{\phi}(\bm{x},\bm{v}), \nabla f(\bm{x},\bm{v})\rangle\d\pi(\bm{x})\d\mu(\bm{v})\\\!=\! 
  \int_{\partial\mathrlap{(\Omega\times\mathcal{V})}} \d\pi(\bm{x})\d\mu(\bm{v})f(\bm{x},\bm{v})  \langle n(\bm{x},\bm{v}), \bm{\phi}(\bm{x},\bm{v}) \rangle   \\
  \!-\!  \int_{\Omega \mathrlap{\times\mathcal{V}}}  \d \pi(\bm{x})\d\mu(\bm{v}) f(\bm{x},\bm{v})
  \Big[ \langle \nabla, \bm{\phi}(\bm{x},\bm{v})\rangle
\\ +\langle \bm{\phi}(\bm{x},\bm{v}), \nabla \ln(\pi(\bm{x})\mu(\bm{v})) \rangle
  \Big].
\end{multline}
As
$\bm{\phi}(\bm{x},\bm{v}) = (\bm{\phi}_x(\bm{x},\bm{v}),
\bm{\phi}_v(\bm{x},\bm{v}))$ , it leads to the invariance conditions,
for $(\bm{x},\bm{v})\in\Omega\times\mathcal{V}$,
\begin{multline}\label{eq:cond-inv}
    \Big (\langle \nabla_x, \bm{\phi}_x(\bm{x},\bm{v}) \rangle\!+\!\langle \nabla_v, \bm{\phi}_v(\bm{x},\bm{v}) \rangle 
    \!+\! \langle \bm{\phi}_x(\bm{x},\bm{v}), -\nabla U(\bm{x})\rangle \\\!+\! \langle \bm{\phi}_v(\bm{x},\bm{v}),\nabla\ln\mu(\bm{v})\rangle  
  \!+\!  \tilde{\lambda}(\bm{x},\bm{v})\Big)\mu(\bm{v})\\ = \int_{\mathrlap{\Omega\times\mathcal{V}}} \tilde{\lambda}(\bm{x},\bm{v}')\tilde{Q}((\bm{x},\bm{v}'), \bm{v})\d\mu(\bm{v}') 
\end{multline}
and, for $(\bm{x},\bm{v})\in\partial(\Omega\times\mathcal{V})$,
\begin{multline}\label{eq:cond-inv-b}
    \langle n(\bm{x},\bm{v}), \bm{\phi}(\bm{x},\bm{v})\rangle_- \mu(\bm{v})\\\!=\! 
    \int_{\mathrlap{\partial(\Omega\times\mathcal{V})}}
           \tilde{Q}_b((\bm{x},\bm{v}'),\bm{v}) \langle n(\bm{x},\bm{v}'), \bm{\phi}(\bm{x},\bm{v}')\rangle_+\d \mu(\bm{v}'),
           \end{multline}

         {\noindent \bf Necessary symmetries.} Thus, considering a
         more general flow $\bm{\phi}$ than the translation
         $\bm{\phi}^T$ a priori imposes different necessary conditions
         from the ones derived in Sec.~\ref{sec:ecmc}. This a priori leads
         to different rates $\tilde{\lambda}$ and event kernel
         $\tilde{Q}$ from the ones derived in
         Section~\ref{sec:necessary_trans}, in order to maintain the
         correct target distribution $\pi\otimes\mu$ invariant. The
         condition on the boundary (\ref{eq:cond-inv-b}) is however
         not impacted, e.g. the boundary kernel $\tilde{Q}_b$ obey the
         condition (\ref{eq:inv_gen_dev-b}) but for a generalized flow
         and can be set to a valid choice $Q$ from the translational
         case.

       First, looking at the conservation condition as obtained in
       (\ref{eq:conservation}), it is now updated to,       
\begin{equation}
  \def\arraystretch{1.4}
  \left\{  \begin{array}{l}
    \int_{\mathcal{V}}  \Big(\langle \nabla, \bm{\phi}(\bm{x},\bm{v}) \rangle+\langle \bm{\phi}_x(\bm{x},\bm{v}), -\nabla U(\bm{x})\rangle \\
    \qquad \qquad\quad+ \langle \bm{\phi}_v(\bm{x},\bm{v}), \nabla \ln\mu(\bm{v})\rangle\Big) \d\mu(\bm{v}) 
             = 0\\
             \int_{\mathcal{V}}  \langle \bm{\phi}(\bm{x},\bm{v}), n(\bm{x},\bm{v})\rangle \d\mu(\bm{v}) 
             = 0\end{array}\right.,
  \label{eq:conservation_gen}
\end{equation}
and the positivity condition imposing the minimal value of the rates
$\tilde{\lambda}$, as obtained in (\ref{eq:min_rate}),  is now updated
to,
\begin{multline}
    \tilde{\lambda}_M(\bm{x},\bm{v}) \!=\! \Big[\langle \bm{\phi}_x(\bm{v}), \nabla U(\bm{x})\rangle \!-\!\langle \nabla, \bm{\phi}(\bm{x},\bm{v}) \rangle \\+ \langle \bm{\phi}_v(\bm{x},\bm{v}), \!-\!\nabla\ln\mu(\bm{v})\rangle\Big]_+.
    \label{eq:min_rate_gen}
  \end{multline}
  As for the translational-flow case, the minimum event rate
    identifies with the rejection rate of the equivalent infinitesimal
    lifted Metropolis scheme. The divergence term
    $-\!\langle \nabla, \bm{\phi}(\bm{x},\bm{v}) \rangle +$ stems from
    the proposal probability ratio \cite{Hastings_1970}. Furthermore, as for the
  translational case, the necessary conditions are imposing the same
  structure for the choice of rates $\tilde{\lambda}$ and kernel
  $\tilde{Q}$, as detailed in (\ref{eq:balance_phi}),
  (\ref{eq:balance_a}) and (\ref{eq:Q_structure}), once updated to the
  minimal-rate condition (\ref{eq:min_rate_gen}). As done for the
  translational case, we now discuss more restrictive but explicit
  classes of valid flows, which obey some further symmetries than the
  necessary ones.
  \\
  
  {\noindent \bf Ideal flows.} Interestingly, it is possible to decrease
  the event rate minimal value to $0$, i.e. to design a flow such that no
  events are needed to target the correct invariance distribution. A
  null minimal rate is equivalent to,
\[\begin{split}
  \langle \nabla, \bm{\phi}(\bm{x},\bm{v}) \rangle \!\geq\!\ 
  \langle \bm{\phi}_x(\bm{v}), \nabla U(\bm{x})\rangle +
  \langle \bm{\phi}_v(\bm{x},\bm{v}),
  \!-\!\nabla\ln\mu(\bm{v})\rangle.\end{split}\] Once combined with the
conservation condition (\ref{eq:conservation_gen}), the inequality is
necessarily tight, so that, 
\begin{equation}
  \def\arraystretch{1.4}
  \begin{array}{l}
    \langle \nabla, \bm{\phi}(\bm{x},\bm{v}) \rangle  \!=\! \langle \bm{\phi}_x(\bm{v}), \nabla U(\bm{x})\rangle + \langle \bm{\phi}_v(\bm{x},\bm{v}), \!-\!\nabla\ln\mu(\bm{v})\rangle,
  \end{array}
  \label{eq:no_event}
  \end{equation}
  which can be summarized as,
   \begin{equation}
     \langle \nabla, \bm{\phi} \rangle \!=\! \langle \nabla \mathcal{H} , \bm{\phi} \rangle.
    \label{eq:no_event_sum}
   \end{equation}
   with
   $\mathcal{H}(\bm{x},\bm{v}) = (U(\bm{x}) - \ln
   \mu(\bm{v}))$, the Hamiltonian of the extended system. It is
   analogous to a continuity equation and the incompressibility
   nature of the probability current $e^{-\mathcal{H}}\bm{\phi}$, leaving
   the Boltzmann distribution stationary. Any \emph{ideal}
   flow then presents a compensation between an increase/decrease of
   the energy of the extended state along $\nabla\mathcal{H}$ and an
   expansion/compression of the infinitesimal volume element, in order
   to preserve the total probability to be in this volume element.

   A particular case then is the one of ideal solenoidal flows which are
   volume-preserving on top of being probability-preserving. They then
   obey the following condition,
\begin{equation}
   \langle \nabla, \bm{\phi} \rangle \!=\! 0  \!=\!   \langle \nabla \mathcal{H} , \bm{\phi} \rangle.
    \label{eq:hamiltonian}
  \end{equation}
  Naturally, setting $\bm{\phi}$ as the Hamiltonian flow
  $\bm{\phi}^H = (\nabla_v\mathcal{H},-\nabla_x\mathcal{H})$ appears
  as a sufficient choice to ensure the correct Boltzmann invariant
  distribution with a null event rate, but its implementation is not
  tractable in almost all cases of interest. More generally,
  Hamiltonian flows belong to a larger class of solenoidal flows,
  which can be characterized as,  
  \begin{equation}
      \bm{\phi}(\bm{x},\bm{v}) = \bm{A}\nabla \mathcal{H},
  \end{equation}
  with $\bm{A}$ some skew-symmetric matrix.
  \\

  {\noindent \bf Uniform-ideal flows.} In the case of a non-ideal flow,
  sampling the Poisson process with minimal event rate
  (\ref{eq:min_rate_gen}) may imply more involved computations than in
  the translational case (\ref{eq:min_rate}). Computations may be eased
  by considering different degree of superposition, e.g.,
  \begin{equation}
    \def\arraystretch{1.4}
    \begin{array}{ll}
(a)        & \langle \nabla(\ln(\pi(\bm{x})\mu(\bm{v})) + \nabla, \!-\!\bm{\phi}(\bm{x},\bm{v}) \rangle_+ \\
 (b) & \langle \nabla(\ln(\pi(\bm{x})\mu(\bm{v})),\!-\!\bm{\phi}(\bm{x},\bm{v}) \rangle_+  \!+\! \langle\nabla, \!-\!\bm{\phi}(\bm{x},\bm{v}) \rangle_+\\
      (c) & \langle \nabla\ln\pi(\bm{x}),\!-\!\bm{\phi}(\bm{x},\bm{v}) \rangle_+ \!+\!\langle \nabla\ln\mu(\bm{v}),\!-\!\bm{\phi}(\bm{x},\bm{v}) \rangle_+\\
      &\qquad\qquad\qquad\qquad\qquad\quad\!+\! \langle\nabla, \!-\!\bm{\phi}(\bm{x},\bm{v}) \rangle_+\\
                       (d)&\dots ,
    \end{array}
    \label{eq:fact_gen}
  \end{equation}
  and implementing a thinning procedure on the different terms.

  These different factorization schemes (\ref{eq:fact_gen}) may indeed
  ease the numerical implementations. However, it may be more advantageous to
  introduce a generalized flow  which reduces the averaged number of
  events, potentially down to the ideal case. However, any flow choice involving a
  precise knowledge of $\nabla U$ will likely be intractable, except
  in some particular simple cases.  Then, a trade-off lies in
  designing flows which do not increase the event rate past its
  physical component (\ref{eq:min_rate}), which is null for instance
  in presence of only hardcore interactions. We then call such flows
  \emph{uniform-ideal} and they must obey the following condition for
  any $(\bm{x},\bm{v})$,
  \begin{equation}
    \langle \nabla, \bm{\phi}(\bm{x},\bm{v}) \rangle + \langle \nabla_v\ln\mu(\bm{v}), \bm{\phi}_v(\bm{x},\bm{v}) \rangle =0.
    \label{eq:flow_phys_ev}
  \end{equation}
  It indeed identifies with the condition (\ref{eq:no_event}) for
  uniform $\pi$ as in hardcore models, making such flow ideal in this
  case. Any ECMC with such generalized flow then comes down to the
  well known case of translational flow, as the conditions of flow
  conservation (\ref{eq:conservation_gen}) and minimal event rate
  (\ref{eq:min_rate_gen}) now identify with the translational ones,
  respectively (\ref{eq:conservation}) and (\ref{eq:min_rate}), so
  that one can respectively set $\tilde{\lambda}$ and $\tilde{Q}$
  to $\lambda$ and $Q$ as obtained in the translational case.

  Similarly to the ideal case, solenoidal flows as,
  \begin{equation}
      \bm{\phi}(\bm{x},\bm{v}) = \bm{A}(\nabla_xf(\bm{x}),\nabla_v\ln\mu(\bm{v})),
  \end{equation}
  with $\bm{A}$ some skew-symmetric matrix and
  $f:\mathcal{S}\to\mathcal{S}$ some function obeys
  (\ref{eq:flow_phys_ev}). It identifies with a rotation in the case
  of $\mu$ Gaussian and $f(\bm{x})=\bm{x}^2$. In the case of a uniform
  distribution $\mu$, any solenoidal flow is valid, which includes
  rotational flows. \\

  {\noindent \bf Hybrid flows.} The PDMP framework allows for a great
  flexibility. Thus, a direct generalization consists in implementing
  sequentially different flows, as a translational flow and a
  rotational one, thanks to an additional discrete lifting
    variable indicating the considered flow (and corresponding event
  rate $\tilde{\lambda}$ and Markov kernel $\tilde{Q}$ if
  necessary). Hybrid flows can be more generally understood as a
    particular case of generalized flows, where the lifting variable
    $v$ admits at least one discrete component which is left invariant
    during the transport. This discrete lifting component, and
  hence the actual implemented flow, can then be resampled at the
  events or refreshments following its stationary distribution
    and the required conditions (\ref{eq:cond-inv}) and
    (\ref{eq:cond-inv-b}). This is most easily done when considering a
    mixture of ideal flows and particularly simplified when
    considering one of uniform-ideal flows. A particular case is when
    it is possible to deterministically switch from one regime to
    another at event (it is always possible at refreshment, given the
    time spent in each regime follows the stationary distribution of
    the lifting variable). This is possible if there is a balance of
    probability flows between the two regimes, i.e., in the case of
    uniform-ideal flows out of simplicity, the generalisation being
    straightforward,
\begin{multline}\label{eq:det-switch}
    \langle \bm{\phi}_x(\bm{x},(\bm{e},k)),\nabla U(\bm{x})\rangle_- \mu((\bm{e},k)) = \\
    \int_{\mathrlap{\Omega\times\mathcal{V}}} \1_{\{k'\}}(k'') \langle \bm{\phi}_x(\bm{x},(\bm{e}',k'')),\nabla U(\bm{x})\rangle_+\\\times Q((\bm{x},(\bm{e}',k'')), \bm{e})\d\mu((\bm{e}',k'')), 
  \end{multline}
where the lifting variable is decomposed along a continuous component $\bm{e}$ and a discrete one $k$.

For instance, the standard ECMC scheme in particle systems can be
understood as some hybrid flows where the discrete lifting component
is the moving particle label. The deterministic switch between the
particle labels involved in the event is possible as
(\ref{eq:det-switch}) is achieved (e.g.
$ \langle \bm{e}, \bm{n}_{ij}\rangle_- = \langle \bm{e},
\bm{n}_{ji}\rangle_+$ for a straight kernel $Q^S$) when considering
the factorized scheme and using the pairwise mirror symmetry. In
contrast, it is not possible to do such a deterministic switch from a
$+x$ to a $+y$ direction, as the necessary flow balance is not
preesnt. A Metropolis-type kernel could be introduced
instead. Similarly, a deterministic switch between two colliding
dimers while generating fixed-radius rotations is not
possible. Therefore generating rotations in such systems, without
backtracking moves, is not trivial. We solve it in
Section~\ref{sec:anisotropy} and in such a way that a deterministic
switch is possible at events (or refreshment) even for the hybrid
flows composed of translations and rotations in the dimer systems, as
implemented in Section~\ref{sec:dimer}.
 
  \section{ECMC with Rotational flow}
  \label{sec:rot}
  As rotational flows constitute a large class of valid uniform-ideal
  flows, we describe in the following their implementations in
  hard-sphere and hard-dimer systems. The generalization to soft
  interactions is straightforward, as rotational flows,
  being uniform-ideal, can directly be implemented in the
  presence of interactions or an external potential by replacing the
  null event rate $\lambda$ to
  $\langle \bm{\phi}(\bm{x},\bm{v}),\nabla U(\bm{x})\rangle_+$.
  
\subsection{Isotropic particle systems: Sphere systems}

Instead of the translational flow as described in
Section~\ref{sec:ecmc_part} parameterized by a lifting variable of the
form $(\bm{e},i)\in\mathbb{S}^1\times\{1,\mydots,N\}$, we now consider
rotational ones, parameterized by the lifting variable
$\bm{v}=(\bm{e},
l,i)\in\mathbb{S}^1\times\mathbb{R}\times\{1,\mydots,N\}$. As for
translations, $\bm{e}$ corresponds to the infinitesimal update applied
to $\bm{x}_i$ but which adds up to a rotation around a center located at 
 $\bm{x}_i + |l|A\bm{e}$ with the infinitesimal rotation matrix
 $A\!=\!\left(\begin{smallmatrix}0&-1\\1&0\end{smallmatrix}\right)$.

 As rotational flows are uniform-ideal,
we can use the choice for $\lambda$ and $Q_b$ described in
Section~\ref{sec:ecmc_part}. Therefore we set,
\begin{equation}
  \def\arraystretch{1.4}
  \left\{
  \begin{array}{l}
    \bm{\phi}_x(\bm{v}) = (0,\dots,\underbrace{\bm{e}}_{i\text{-th}},\dots,0)\\    
    \bm{\phi}_v(\bm{x},\bm{v})=(\tfrac{1}{l}A\bm{e},0,0)\\
    \lambda(\bm{x},\bm{\theta},\bm{v})=0\\
    \mu(\bm{e},l,i)= \tfrac{1}{2\pi}\1_{\mathbb{S}^1}(\bm{e})\nu(l)\frac{1}{N}\1_{\{1,\dots,N\}}(i)\\
    \end{array}\right. .
  \label{eq:flow_sphere_algo}
\end{equation}
 The
  corresponding algorithm corresponds to the Algorithm~\ref{algo:ECMC}
  with those choices. Such flow is indeed uniform-ideal, as
$\langle \nabla, \bm{\phi}\rangle=0$. Apart no positive mass on
  $l=0$ out of simplicity, it is noteworthy that there is no
constraint on the distribution of the auxiliary variable $l$, setting
the distance between the rotation center and $\bm{x}_i$ but also the
anticlockwise or clockwise nature of the rotation. Thus, for
  instance, one can set $\nu(l)$ so that the generated rotations are
always clockwise and at a fixed distance $|l|>0$ or so that
  $\nu(l)$ is a continuous distribution, e.g. a Gaussian. Such
flexibility then offers many possibilities to optimize the flow to a
given problem. The generalisation to the case of $l$ admitting a
  positive mass on $0$, which is equivalent to introducing stalling
  time on every configuration, is straightforward by including to the
  distribution of $\bm{e}$ a positive mass on $0$ \footnote{In
      more details, the changes are
      $\bm{\phi}_v(\bm{x},\bm{v})=\1_{\mathbb{R}^*}(l)(\tfrac{1}{l}A\bm{e},0,0)$
      and
      $\mu(\bm{e},l,i)=
      \big(\tfrac{\nu_0}{1+\nu_0}\1_{\{0\}}(l)1_{\{0\}}(\bm{e}) +
      \tfrac{1}{1+\nu_0}\1_{\mathbb{R}^*}(l)\tfrac{1}{2\pi}\1_{\mathbb{S}^1}(\bm{e})\nu(l)\big)\frac{1}{N}\1_{\{1,\dots,N\}}(i)$.}

Regarding the kernel $Q_b$, as was previously mentioned, it can be set
to a deterministic kernel (\ref{eq:straight_kernel}) or to the direct
one (\ref{eq:direct_kernel}). As the flow $\bm{\phi}_x$ is directly
parameterized in terms of an infinitesimal vector $\bm{e}$, the
implementation is straightforward as it comes down to resampling
$\bm{e}$, which then directly resample the rotation center position
$\bm{x}_i + |l|A\bm{e}$. Here again the variable $l$ does not
  have an impact in the event and can be left invariant or resampled
  so as to leave $\nu(\cdot)$ invariant.

Finally, we study the numerical performances of implementing
non-reversible rotations in sphere systems in
Section~\ref{sec:spheres}.

\subsection{Anisotropic particle systems: Dimer systems}
\label{sec:anisotropy}

The hard-dimer system, modeled closely after hard spheres, provides a
simple setting to study the rich behavior and the phase transitions of
anisotropic particles \cite{Wojciechowski_1993,Cugliandolo_2017}. Such
particles then require some degree of rotation and we now devise a
ECMC schemes generating them in a completely non-reversible manner.
\\

{\bf\noindent Dimer configurations.} We now consider a system of $N$
hardcore dimers of radius $\sigma$ in a periodic $L$-box.
Their configurations are described by the following state variable,
$(\mathbf{x},\bm{\theta})\in (\mathbb{R}/(L\mathbb{Z}))^{2N}\times
(\mathbb{R}/(\pi\mathbb{Z}))^N$, where $\bm{x}$ is the sequence of
dimer centers and $\bm{\theta}$ the one of angles. We furthermore
introduce the following notations for the position of the two monomers
constituting one dimer,
$$\bm{x}_i^+= \bm{x}_i + \sigma \bm{\delta}_\parallel(\theta_i) , \bm{x}_i^{-}= \bm{x}_i - \sigma \bm{\delta}_\parallel(\theta_i).$$ with $\bm{\delta}_\parallel(\theta) = (\cos(\theta),\sin(\theta))\in\mathbb{S}^1$
for $\theta\in[0,\pi]$.
The valid configurations forming up the set $\Omega$ satisfy the
following constraint for every pair of dimers,
\begin{equation}
  d\left(\bm{x}^\pm_i , \bm{x}_j^\pm\right) > d_{\text{pair}}=2\sigma,
  \label{eq:constraint}
\end{equation}

Now, similarly to the sphere case, in order to devise an ECMC
sampling, we first extend the configurations to
$(\bm{x},\bm{\theta},\bm{v})$ with
$\bm{v}=(\bm{e}_+,l_\perp,i)\in\mathbb{S}^1\times\mathbb{R}\times\{1,\dots,N\}$
where $i$ is the label of the updated dimer, $\bm{e}_+$ the update
vector of the $+$-monomer and $l_\perp$ is the variable setting,
respectively to $\bm{e}_+$ and $\bm{x}^+_i,\bm{x}^-_i$, the update
vector $\bm{e}_-$ of the $-$-monomer, so that (\ref{eq:constraint}) is
obeyed, i.e.,
\begin{equation}
  \bm{e}^- = \bm{e}^+ +l_\perp\delta_\perp(\theta_i),\ \text{with}\  \delta_\perp(\theta)=(-\sin\theta,\cos\theta).
  \label{eq:epm}
  \end{equation}
  It is naturally possible to devise a scheme directly at the dimer
  level, updating $\bm{x}_i$ and $\theta_i$. However, introducing the
  infinitesimal update of each monomer enables an easier handling of
  the dimer constraint and facilitates the direct utilization of the
  kernels devised in the translational case in
  Section~\ref{sec:ecmc_part}.  Furthermore, contrary to sphere
  systems, the dimer constraint imposes a correlated update of each of
  its monomer and it leads to requirements on the choice of
  $l_\perp$. For instance, setting $l_\perp$ to some non-zero fixed
  value cannot give a valid scheme, as mentioned earlier. Indeed each
  dimer could only rotate in a single manner, as enforced by
  (\ref{eq:epm}). Then flows are not conserved and
  (\ref{eq:conservation_gen}) not satisfied, as there is no label switch
  symmetry. The scheme could be correct at the cost of introducing
  both clockwise and anticlockwise rotations and proposing, at
  collision, to backtrack the rotation of the updated dimer. Therefore
  we now derive the required distribution of $l_\perp$ in order to
  ensure a completely non-reversible process.

As it is necessary for
ergodicity to rotate the dimers, the monomer update vectors $\bm{e}_+$
and $\bm{e}_-$ should cover the whole hypersphere $\mathbb{S}^1$ and
therefore we choose for $\bm{e}_+$, and by symmetry for $\bm{e}_-$,
the uniform distribution on $\mathbb{S}_1$, another possible choice
being a Gaussian distribution for instance. Doing so, it constrains
the possible choice of $l_\perp$ to $0$ (i.e. translation) or to a
value depending on $\theta_i$, which we obtain, parameterizing
for the $i$-th dimer,
\begin{equation}
  \bm{e}_+ = (\cos(\theta_i+\alpha),\sin(\theta_i+\alpha)), \alpha\in[0,2\pi],
  \label{eq:e+}
  \end{equation}
as,
\begin{equation}
\bm{e}_- =  (\cos(\theta_i-\alpha),\sin(\theta_i-\alpha)), l_\perp = 2\sin \alpha.
  \label{eq:e-}
  \end{equation}
  Thus, any normed choice $(\bm{e}_+,\bm{e}_-)$ corresponds to a
  rotation of the dimer around a center placed on the dimer bisector,
  hence the derivation of what we call \emph{bisector} rotational
  flow, as now detailed and illustrated on Fig.~\ref{fig:schemes}.
  \\

  {\noindent \bf Bisector rotational flow.} We now formalize the
    flow for a rotation of a given dimer (i.e. of both its monomers)
    around a center placed on its bisector at
    $\bm{x}_i + \tan\alpha\bm{\delta}_\perp(\theta_i)$. Out of simplicity, we
  now set $\bm{v}=(\alpha,i)\in[0,2\pi]\times\{1,\dots,N\}$, so that
  the ECMC sampling by bisector rotational flow is completely
  characterized by,
\begin{equation}
  \def\arraystretch{1.4}
  \left\{
  \begin{array}{l}
    \bm{\phi}_x(\bm{\theta},\bm{v}) = (0,\dots,\underbrace{\cos\alpha\bm{\delta}_\parallel(\theta_i)}_{i\text{-th}},\dots,0)\\    
   \bm{\phi}_\theta(\bm{v})\! =\! (0,\dots,\underbrace{\sin\alpha/\sigma}_{i\text{-th}},\dots,0)\\
    \bm{\phi}_v(\bm{x},\bm{v})=0\\
    \lambda(\bm{x},\bm{\theta},\bm{v})=0\\
    \mu(\alpha,i)= \frac{1}{2\pi}\1_{[0,2\pi]}(\alpha)\frac{1}{N}\1_{\{1,\dots,N\}}(i)\\
    \end{array}\right. ,
  \label{eq:flow_dimer}
\end{equation}
and any choice of $Q_b$ corresponding to a valid one for sphere
systems, up to the mapping of $\alpha$ to $\bm{e}_+$ in (\ref{eq:e+})
or $\bm{e}_-$ in (\ref{eq:e-}), depending on which monomer is involved
in a collision, and then mapping back from the resampled $\bm{e}'_\pm$ to
$\alpha'$.

From an algorithmic point of view and collision computations, it may
however be easier to work with a parametrization directly based on the
rotation center. We then set
$\bm{v}=(l, \gamma,i)\in\mathbb{R}\times\{-1,1\}\times\{1,\dots,N\}$
to fix the rotation center ($\bm{x}_i + l\bm{\delta}_\perp(\theta_i)$)
and sign ($\gamma=+1,-1$ for respectively anticlockwise, clockwise), so that,
\begin{equation}
  \def\arraystretch{1.4}
  \left\{
  \begin{array}{l}
    \bm{\phi}_x(\bm{\theta},\bm{v}) = (0,\dots,\underbrace{\tfrac{\gamma l}{\sqrt{l^2+\sigma^2}}\bm{\delta}_\parallel(\theta_i)}_{i\text{-th}},\dots,0)\\    
   \bm{\phi}_\theta(\bm{v})\! =\! (0,\dots,\underbrace{\tfrac{\gamma}{\sqrt{l^2+\sigma^2}}}_{i\text{-th}},\dots,0)\\
    \bm{\phi}_v(\bm{x},\bm{v})=0\\
    \lambda(\bm{x},\bm{\theta},\bm{v})=0\\
    \mu(l,\gamma,i)= \frac{1}{\pi}\frac{\sigma}{l^2+\sigma^2}\frac{1}{2}\1_{\{-1,1\}}(\gamma)\frac{1}{N}\1_{\{1,\dots,N\}}(i)\\
    \end{array}\right. ,
  \label{eq:flow_dimer_algo}
\end{equation}
leading to the following mapping to $\bm{e}_\pm$ for the $i$-th dimer,
\begin{equation}
  \left\{
    \def\arraystretch{1.4}
    \begin{array}{l}
             \bm{e}_+= \tfrac{\gamma}{\sqrt{l^2+\sigma^2}}(l\bm{\delta}_\parallel(\theta_i)+\sigma\bm{\delta}_\perp(\theta_i))\\
             \bm{e}_-=\tfrac{\gamma}{\sqrt{l^2+\sigma^2}}(l\bm{\delta}_\parallel(\theta_i)-\sigma\bm{\delta}_\perp(\theta_i))
    \end{array}\right..
\label{eq:mapping_e}
\end{equation}
Such mapping can be used to implement $Q_b$, as previously
mentioned. First, the vector $\bm{e}_+$ or $\bm{e}_-$, depending on
which monomer is involved in a collision, is reampled. Then the
resampled values $l',\gamma',i'$ are obtained using
(\ref{eq:mapping_e}). For both parametrization, checking that the flow
is uniform-ideal ($\langle\nabla,\bm{\phi}\rangle=0$) is
straightforward and the corresponding algorithms correspond to the
Algorithm~\ref{algo:ECMC} with those choices.

As can be seen from (\ref{eq:flow_dimer_algo}), the correct
distribution of rotation center is not trivial and especially not
uniform in space. Indeed, the distance $l$ between the dimer and
rotation centers actually follows a Cauchy distribution. High
values for $l$ are not rare and should be treated numerically with
care regarding float precision. In particular, the situation $l=0$ correspond to self-rotations,
while the value $l\to \pm\infty$ codes for translations. Thus all
possible single-dimer rotations with normed monomer velocities can
indeed be generated.

We study the numerical performances of implementing non-reversible
rotations in dimer systems in Section~\ref{sec:dimer}.

\section{Numerical experiments}
\label{sec:numerics}
\subsection{Rotational flows in hard-sphere systems}
\label{sec:spheres}
\newcommand{\straightTXY}{StraightTXY\xspace}
\newcommand{\directT}{DirectT\xspace}
\newcommand{\straightR}{StraightR\xspace}
\newcommand{\directR}{DirectR\xspace}

\begin{figure}
 \includegraphics{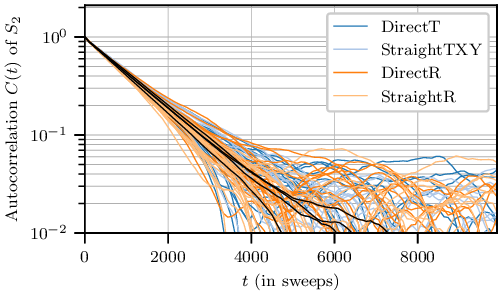}
 \caption{Autocorrelation of the parameter $\Psi_6$, for 20 runs and
   their averaged autocorrelation (black). The system consists of $N=64$
   spheres at density $\rho=0.708$. One sweep 
   corresponds to $N$ event computations. The \directT, \directR and \straightR schemes are with no refreshment, the \straightTXY is at a tuned refreshment rate of $d_{\text{ref}}/L=0.198$.}
\label{fig:autocorr_sphere}
\end{figure}

\begin{figure}
\includegraphics[width=0.49\textwidth]{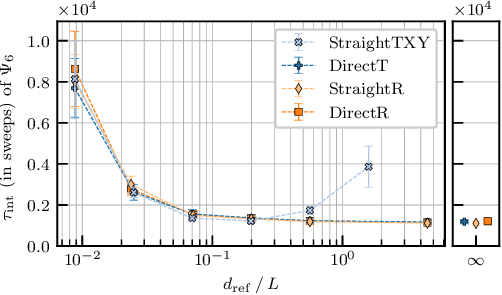}\hfill
\includegraphics[width=0.49\textwidth]{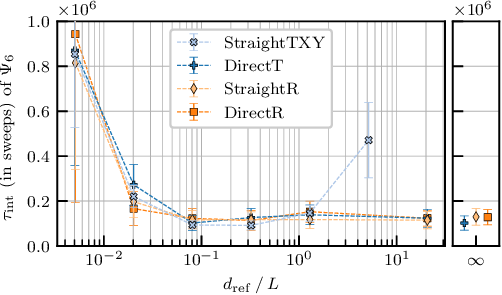}
\caption{$\Psi_6$ integrated autocorrelation time as a function of the
  refreshment distance $d_{\mathrm{ref}}$, for $N=64$ ({\bf left}) and
  $N=256$ ({\bf right}) hard spheres at density
  $\rho=0.708$. Rotations have a radius $\ell=4\sigma$. One sweep
  corresponds to $N$ event computations and $L$ is the size of the
  system.}
\label{fig:sphere_tau_vs_ref}
\end{figure}

We first study rotational flows for isotropic hard spheres, as
presented in Section \ref{sec:flow}. We consider a system of $N$
spheres of radius $\sigma$ in a periodic box of size $L$ at the
vicinity of the liquid-hexatic phase transition \cite{Bernard_2011},
setting $\rho = N \pi \sigma^2 / L^2= 0.708$. In order to estimate the 
relaxation time, we consider the decorrelation of the a priori slowest
observable, here being $\Psi_6$ the global orientational
parameter \cite{Strandburg_1988}. It is defined as the average over
each sphere $j$ of the average of the angles of the bonds $\phi_{jk}$
with its $n_j$ nearest neighboring particles, i.e.,
\begin{equation} \Psi_6 = \frac 1 N \sum_{j=1}^N \frac{1}{n_j}
  \sum_{k=1}^{n_j} \exp(6i\phi_{jk}) . \label{eq:Psi_6} \end{equation}

We compare two translational schemes, respectively along $+x,y$ with a
straight kernel (\straightTXY) and along all directions in
$\mathbb{S}^1$ with a direct kernel (\directT), and two rotational
schemes generating all clockwise rotations of radius $\ell=4\sigma$,
respectively with a straight kernel (\straightR) and a direct one
(\directR). At fixed-time refreshments, all of the lifting variables
are drawn again according to their invariant measure. As all schemes
have constant velocity, it corresponds to a fixed refreshment distance
$d_{\mathrm{ref}}$. Agreement between algorithms can
  be checked for the pair correlation function in
  Appendix~\ref{app:pair_corr}.

First, as shown in Fig.~\ref{fig:autocorr_sphere}, the
  autocorrelation function for $\Psi_6$ appears close to an
  exponential decay, and in particular does not exhibit different time
  scales. Therefore, the integrated autocorrelation time
  $\tau_{\text{int}}$ is a well-suited measure of both the
  decorrelation time and effective sample size. We compute an averaged
  $\langle\tau_{\text{int}}\rangle$ over 20 independent runs and
  derive the error bars as being 3 times the standard deviation on
  these 20 runs. On each run, $\tau_{\text{int}}$ is obtained by
  summing the autocorrelation function from $0$ to some
  $k_{\text{max}}$ value, belonging to the plateau regime of $\tau_{\text{int}}$ vs
  $k_{\text{max}}$. We set the unit of time to be equal to one sweep (i.e. $N$
  events), as the computational complexity is the same for all schemes
  and such time unit is independent from code details.

\begin{figure}
\includegraphics{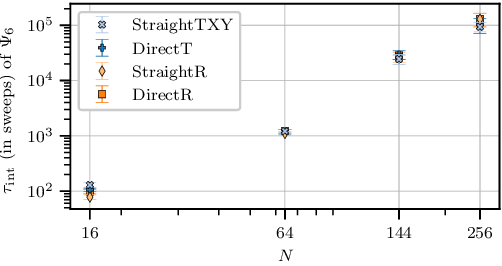}
\caption{$\Psi_6$ integrated autocorrelation time as a function of the 
number of particles $N$, for hard spheres at density $\rho=0.708$. The 
\straightTXY has optimal refreshment, the other schemes have no refreshment. 
Rotations have a radius $\ell=4\sigma$. One sweep corresponds to $N$ event 
computations.}
\label{fig:sphere_tau_vs_N}
\end{figure}

Now, it clearly appears on
  Fig.~\ref{fig:sphere_tau_vs_ref} that the \straightTXY scheme
  requires some fine-tuning in order to set $d_{\mathrm{ref}}$ to an
  optimal value, as already observed \cite{Bernard_2009}. On the
  contrary, all other schemes (\directT, \straightR, and \directR)
  show better decorrelation performance, as the refreshment time
  increases. This was also already observed for some schemes
    \cite{Michel_2020,Hollmer_2022}, it is however a particularly
    interesting behavior for the \straightR scheme. Indeed, as
    $d_{\text{ref}}\to \infty$, it becomes a purely deterministic
    process but apparently it still yields an ergodic and efficient
    scheme.  Then, while the impact of the refreshment time greatly
    depends on the considered scheme, the scaling of the integrated
    autocorrelation time of $\Psi_6$ with the number of particles $N$
    appears similar for all four schemes, as shown in
    Fig.~\ref{fig:sphere_tau_vs_N}.

\begin{figure}
\includegraphics{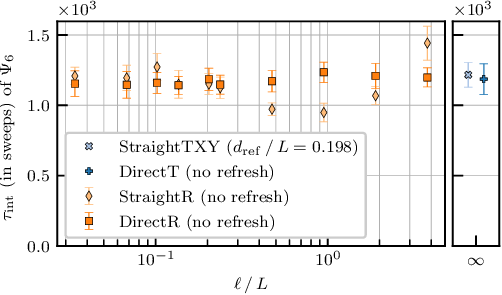}
\caption{$\Psi_6$ integrated autocorrelation time as a function of the radius 
$\ell$, for $N=64$ spheres at density $\rho=0.708$. The fine-tuned versions 
of the translational schemes are shown for reference. 
One sweep corresponds to $N$ event computations and $L$ is the size of the 
system.}
\label{fig:sphere_different_l}
\end{figure}

Finally, regarding the schemes generating rotational flows, we found
that no gain in performance was achieved by introducing switches at
events between clockwise and anti-clockwise rotations. Also,
the choice of the rotation radius $\ell$ appears not critical, as can
be seen on Fig.~\ref{fig:sphere_different_l}, where only the
\straightR scheme shows some $\ell$-dependence and a slightly better
performance than all the other fine-tuned schemes for $\ell \sim L$.

\subsection{Bisector rotational flows in hard-dimer systems}
\label{sec:dimer}

\newcommand{\ECMCmetro}{ECMC+Met\xspace}
\newcommand{\Rbisector}{Rbisector\xspace}
\newcommand{\Alternate}{Hybrid\xspace}

\begin{figure}
 \includegraphics{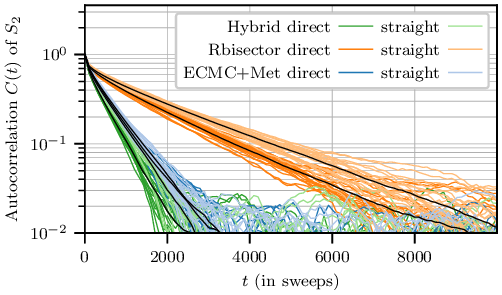}
 \caption{Autocorrelation of the parameter $S_2$, for 20 runs
     and their averaged autocorrelation (black). The system consists
     of $N=32$ dimers at density $\rho=0.7$. The \Alternate and
     \Rbisector schemes do not have any refrehsment and the \ECMCmetro has a
     tuned refreshment rate. One sweep corresponds to $N$ event
     computations.}
\label{fig:autocorr}
\end{figure}

\begin{figure*}
 \includegraphics[width=0.49\textwidth]{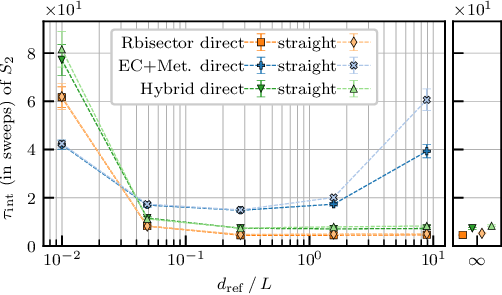}\hfill
 \includegraphics[width=0.49\textwidth]{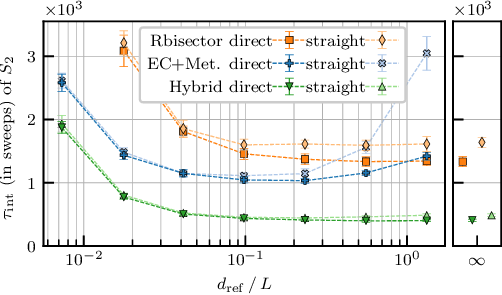}
 \caption{Nematic scalar order parameter
   $S_2 = \left\langle 2 \cos^2(\theta_j - \bar\theta) -1
   \right\rangle$ integrated autocorrelation time as a function of the
   refreshment distance $d_{\mathrm{ref}}$, for $N=32$ hard dimers at
   density $\rho=0.5$ ({\bf left}) and $\rho=0.7$ ({\bf right}). One
   sweep corresponds to $N$ event computations and $L$ is the size of
   the system.}
\label{fig:dimer_tau_scalar_vs_ref}
\end{figure*}

We now study the efficiency of rotational flows in the anisotropic
case of the hard-dimer system, as described in section
\ref{sec:anisotropy}. The system consists in $N=32$ dimers of monomer
radius $\sigma$ in a periodic box of size $L$. Building on previous Monte
Carlo \cite{Wojciechowski_1993} and molecular dynamics
simulations \cite{Cugliandolo_2017}, simulations are performed at
densities $\rho= 2N \pi \sigma^2 / L^2= 0.5$ and $\rho = 0.7$, the
latter being slightly under the observed phase transition. As
commonly studied in isotropic-nematic transition \cite{Eppenga_1984},
we now consider the relaxation of $S_2$, the scalar parameter for the
two-dimensional nematic order,
\begin{equation}
S_2 = \frac 1 N \sum_{j=1}^N \left( 2 \cos^2(\theta_j - \bar\theta) -1 \right) 
\in [0,1],
\end{equation}
where $\bar\theta$ is the angle of the director, defined modulo
$\pi$. This observable has an intrinsic dimer-flip symmetry
$\theta_j \mapsto \theta_j+\pi$. Other observables, with or without
the dimer-flip symmetry, can be constructed with the
$\theta_j$'s. They are presented in Appendix~\ref{app:obs} and show
a consistent behavior. In previous work \cite{Cugliandolo_2017}, the $\Psi_6$ of the
equivalent sphere configuration is investigated, but we found that the
global order of the $\theta_i$'s shows longer correlation times at the
considered densities.

We consider two different algorithms: The \Rbisector scheme with a
bisector rotational flow, as introduced in Sec.~\ref{sec:anisotropy},
and the \Alternate one with a flow switching at event between
translational and bisector rotational flows. Both are studied combined
either with the straight and direct kernels. We compare them to the
\ECMCmetro algorithm consisting in a ECMC scheme combined with
Metropolis rotation proposals, as already investigated for some
anisotropic particle systems\cite{Klement_2021}. Here, the ECMC flow
consists in translational moves, with all directions allowed. At each
refreshment, a random number $n$ of dimer rotations are also
proposed. Their number $n$ follows a geometric distribution of
parameter $p_{\mathrm{chain}}$ and the proposed rotation angle
increments are uniformly picked over $[-\Delta, \Delta]$, $\Delta$
being tuned at $\Delta=\pi$ for $\rho=0.5$ and $\Delta=0.4$ for
$\rho=0.7$.  Following the previous work\cite{Klement_2021},
$p_{\mathrm{chain}}$ is taken so that on average there is as many ECMC
events as there are Metropolis proposals and both kind counts as one
event computation. In addition to the straight kernel, we also
simulate the \ECMCmetro scheme with a direct kernel. At fixed-time
refreshments, all lifting variables are fully drawn again according to
their invariant measure. As all schemes have constant velocity, it
corresponds to a fixed refreshment distance $d_{\mathrm{ref}}$.
Agreement between algorithms can be checked for the
  pair correlation function in Appendix~\ref{app:pair_corr}.

As shown in Fig.~\ref{fig:autocorr}, the autocorrelation function
  for $S_2$ appears close to a single-timescale exponential decay, and
  we apply the same procedure as in Section~\ref{sec:spheres} to
  estimate the integrated autocorrelation time. We also set the time
  unit to be one sweep as the computational complexity is also the
  same for all schemes. Its evolution according to the refreshment is
displayed in Fig.~\ref{fig:dimer_tau_scalar_vs_ref}. For the dilute
case $\rho=0.5$, the \ECMCmetro scheme shows an optimal value at
finite $d_{\mathrm{ref}}$ for both the straight and direct
  kernels, whereas the \Rbisector and \Alternate schemes reach an
optimum for vanishing refreshment $d_{\mathrm{ref}} \to
\infty$. Similarly to the rotations in sphere systems, the \Rbisector
and \Alternate schemes with straight kernel turn into purely
deterministic processes as $d_{\mathrm{ref}} \to \infty$. Their direct
schemes outperform their straight counterparts and the fastest
decorrelation is achieved with the \Rbisector algorithm (speed-up of
$\sim 3.0$ compared to ECMC+Met, the \Alternate achieves a speed-up of
$\sim 2.1$). At the denser density of $\rho = 0.7$, similar
observations can be drawn, regarding the direct versions of the
algorithms better performing and the $d_{\mathrm{ref}}$-tuning
requirement only for the \ECMCmetro algorithms. However, the
\Rbisector scheme now performs worse than the \ECMCmetro scheme
(speed-up of $~\sim 0.74$) and the \Alternate algorithm shows the
fastest decay (speed-up of $\sim 2.6$). Such behavior could be
explained from the observed collision loops in this denser regime,
where a rotating dimer $i$ will collide with another dimer $j$, which
will collide in turn with $i$ and so on. Such loops are of finite
  time but hinders the dynamics. There, introducing a flow switch
between rotations and translations seems to control the appearance of
such loops.

\section{Conclusion}

Since the introduction of the ECMC method, important algorithmic and
analytical efforts have been made towards the generalization of such
schemes to any systems. Drawing on that line, the PDMP
characterization of ECMC schemes allows us to derive, first for
  translational flows, the necessary fundamental symmetries of the
flow conservation and minimum event rate, as required for invariance
of the correct distribution $\pi$. It then makes clearly appear the
more restrictive but sufficient symmetries most commonly imposed in
order to derive explicit schemes. From there, the introduction of
generalized flow is carried on in the same manner, by a careful
characterization of necessary and sufficient symmetries. We
  recover the necessary flow conservation and minimum event rate
  conditions. In particular, we define two classes of flows of
interest: the ideal one, where events are suppressed, and the
uniform-ideal ones, for which one can use the valid choices of rates
and kernels, as developed and known for the usual translational flow.

In particular, the uniform-ideal class includes rotational flows and
we devise ECMC schemes generating completely non-reversible rotations
in bidimensional hard-sphere and hard-dimer systems. While the
rotations can be simulated in many different ways in hard spheres, the
contact constraint between the two monomers forming each individual
dimer imposes strict limitations on feasible rotations. These
limitations ultimately narrow down the potential rotations to the
bisector rotational flow when considering normed monomer
velocities. Importantly, numerical simulations indicate,
at least for meaningful observables and high densities, that for
rotational flows the refreshment mechanism does not require some
crucial fine tuning and may actually not be necessary for ergodicity,
even for schemes then completely deterministic. This is reminiscent of
numerical observations of similar behavior for translational ECMC with
a reflection kernel in hard-sphere systems
\cite{Bernard_2009}. Proving such property constitutes a major
mathematics challenge, as the question of proving ergodicity of ECMC
schemes with refreshment for hard-particle systems is already not
trivial \cite{Monemvassitis_2023} and an ergodicity proof with no
refreshment, but with some conditions on the potential, is only known
for the so-called Zig-Zag process \cite{BRZ19}.  In lack of such
proof, a refreshment distance much larger than $L$ could however be
chosen.

Such possibility of achieving ergodicity without any refreshment
indicates how the inherent chaotic nature of the system can be
harnessed into producing efficient and robust non-reversible
schemes. Comparing the performances observed for the hard-sphere
system and hard-dimer one, the latter presenting an additional nematic
order, showcases how the intrinsic stochastic nature of the system
impacts the optimization of the algorithm scheme. It indeed appears
that hard-sphere system close to the transition point does not
particularly benefit from more involved schemes, apart from the
robustness to the refreshment fine tuning. However, one can note that
the \straightR scheme at $\ell \simeq L$ is better performing, while
actually generating some very gradual sequential update of the
direction. This behavior is reminiscent of sequential translational
straight schemes where the direction is incremented by a fixed value
at refreshments, for example in tethered dimer systems
\cite{Hollmer_2022, Qin_2022}. On the other hand, in the hard-dimer
case, speed up are observed, even for the small considered system
size, and at the dilute and denser densities. Interestingly, at the
denser density, the \Alternate scheme allows to control the collision
loops, which arise from the strong orientational correlation and may
slow down the process when considering only bisector rotations.  Thus,
it appears that developing efficient non-reversible algorithms require
a fine understanding of the correlations present in the system. An
interesting prospect is to consider the implementations of rotational
flows to other anisotropic-particle systems. Understanding the
constraints imposed by particle shape and how anisotropy, alongside
resulting correlations, influences algorithm efficiency could yield
valuable insights.

Certainly, further work should also deal with the efficient numerical
implementation of rotational flows, with in addition soft
interactions. The challenge here lies in the simulations of the event
times, which should computationally benefit from a thinning
procedure. More generally, the efficient parallelization of such
rotational flows is an important question, especially for applying
these methods to large-scale molecular simulations. Efforts towards
parallelization have already been made for standard translational ECMC
\cite{kampmann_2015_para,Li_2021}. The parallelization of rotational
flows could make it possible to study the scaling and evolution of the
observed speed-up in hard-dimer for large system sizes. Additionally,
such parallelization is essential to accurately characterize their
critical behavior.

\begin{acknowledgments}
All the authors are grateful for the support of the French ANR under
the grant ANR-20-CE46-0007 (\emph{SuSa} project).  A.G. is supported by the Institut Universtaire de France. Computations have
been performed on the supercomputer facilities of the Mésocentre
Clermont Auvergne.
\end{acknowledgments}

\section*{Conflict of Interest Statement}

The authors have no conflicts to disclose.

\section*{Data Availability Statement}

The data that support the findings of this study are available from
the corresponding author upon reasonable request. The codes used to
generate the data are available at
\url{https://plmlab.math.cnrs.fr/stoch-algo-phys/ecmc/particle-rotational-flow}.

\bibliography{biblio}

\clearpage

\onecolumngrid

\appendix

% \begin{center}
%   \large{\bf Supplementary Materials\\Necessary and sufficient
%     symmetries in Event-Chain Monte Carlo\\
%     with generalized flows and
%     Application to hard dimers}
% \end{center}

\section{Simulation agreements}
\label{app:pair_corr}

\begin{figure}[ht]
\includegraphics[width=0.49\textwidth]{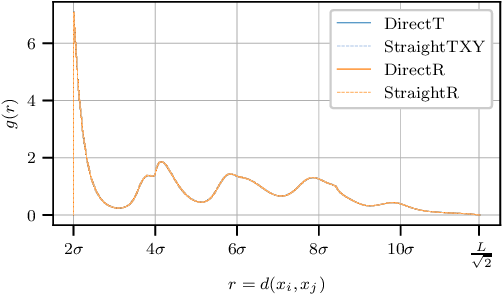}\hfill
\includegraphics[width=0.49\textwidth]{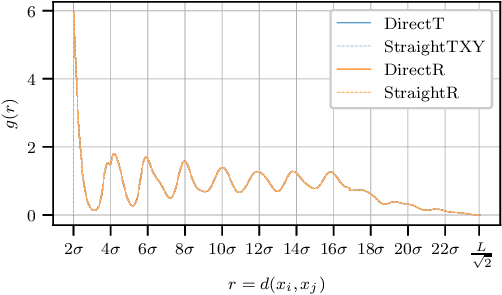}
\caption{Pair correlation $g(r)$ between spheres for a system with
  $N=64$ ({\bf left}) and $N=256$ ({\bf right}) hard spheres with
  density $\rho=0.708$, for the considered variants of ECMC.}
\label{fig:pair_correlation_sphere}
\end{figure}

\begin{figure}[ht]
\includegraphics[width=0.49\textwidth]{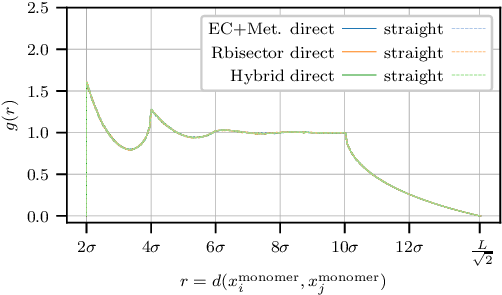}\hfill
\includegraphics[width=0.49\textwidth]{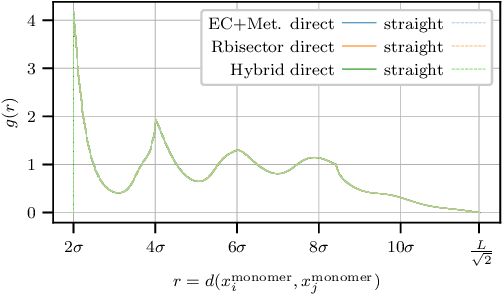}
\caption{Pair correlation $g(r)$ between
  two monomers belonging to different dimers for a system with $N=32$
  hard dimers with density $\rho=0.5$ ({\bf left}) and $\rho=0.7$
  ({\bf right}), for the considered variants of ECMC.}
\label{fig:pair_correlation_dimer}
\end{figure}

Here we provide a comparison of the full pair correlation
function estimated by the algorithms considered for the sphere systems
in Figure~\ref{fig:pair_correlation_sphere} and for the dimer ones in
Figure~\ref{fig:pair_correlation_dimer}.

\section{Additional dimer observables}
\label{app:obs}
The dimer angles 
$\theta_j$ can be seen twofold, either with $\pi$- or $2\pi$-periodicity. If 
opposite 
directions can be discriminated, the global orientational order is encoded in 
the average of the directions $\bm{\delta}_\parallel(\theta_j) = (\cos \theta_j, \sin 
\theta_j)$ of all dimers, namely the polarization,
\begin{equation}
\bm{p} = \frac 1 N \sum_{j=1}^N \bm{\delta}_\parallel(\theta_j) .
\end{equation}

If the system has an intrinsic dimer-flip symmetry
$\theta_j \mapsto \theta_j+\pi$, the relevant observable stems from
the study of the isotropic-nematic phase transition
\cite{Eppenga_1984} in three dimensions. The general matrix order
parameter of such a transition is reduced in dimension $2$ to a
complex number,
\begin{equation}
 z_2 = \frac 1 N \sum_{j=1}^N \exp(2i\theta_j) 
= S_2 \exp(2i \bar\theta) ,
\end{equation}
where $\bar\theta$ is the angle of the director, defined modulo $\pi$, and 
$S_2$ is the scalar parameter for the two-dimensional nematic order,
\begin{equation}
S_2 = \frac 1 N \sum_{j=1}^N \left( 2 \cos^2(\theta_j - \bar\theta) -1 \right) 
\in [0,1].
\end{equation}
In the following, $z_2$ is called the nematic vector. Its expression
is reminiscent of the $\Psi_6$ for hard disks and plays the role of
averaging in a $\pi$-periodic setting.  The square box imposes the
ensemble averages $\langle z_2 \rangle = 0$ and
$\langle \mathbf p \rangle = 0$, which give a rudimentary check for
ergodicity, respectively with and without dimer-flip symmetry.

\begin{figure}[ht]
\includegraphics[width=0.49\textwidth]{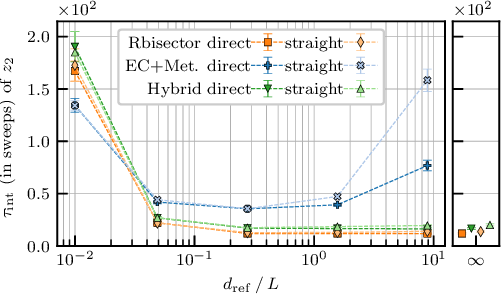}\hfill
\includegraphics[width=0.49\textwidth]{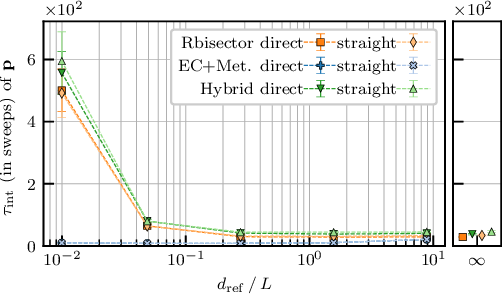}
\caption{Nematic vector $z_2$ (left) and polarization $\mathbf p$ (right) 
integrated autocorrelation time as a function of the refreshment distance 
$d_{\mathrm{ref}}$, for $N=32$ hard dimers at density $\rho=0.5$. One sweep 
corresponds to $N$ event computations and $L$ is the size of the system.}
\label{fig:dimer_tau_vs_ref_liquid}
\end{figure}

\begin{figure}
\includegraphics[width=0.49\textwidth]{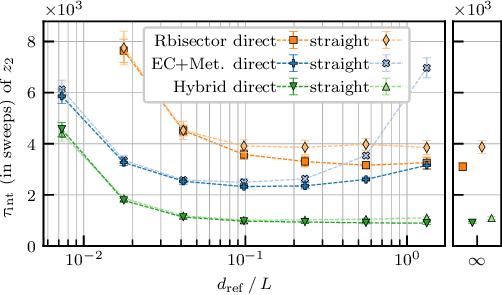}\hfill
\includegraphics[width=0.49\textwidth]{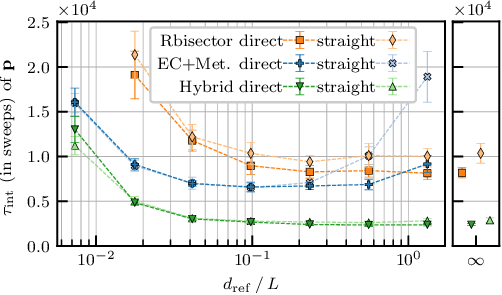}
\caption{Nematic vector $z_2$ (left) and polarization $\mathbf p$ (right) 
integrated autocorrelation time as a function of the refreshment distance 
$d_{\mathrm{ref}}$, for $N=32$ hard dimers at density $\rho=0.7$. One sweep 
corresponds to $N$ event computations and $L$ is the size of the system.}
\label{fig:dimer_tau_vs_ref_dense}
\end{figure}

Experiments (Fig. 
\ref{fig:dimer_tau_vs_ref_liquid},\ref{fig:dimer_tau_vs_ref_dense}) show that the 
polarization is slower than the nematic vector, but the need to flip all dimers 
is an artificial constraint in the case where opposite directions are 
equivalent. The scalar $S_2$ decorrelates faster than $z_2$ but the behavior of 
the algorithms is the same for all observables.

At low density, the \ECMCmetro scheme has an artificial speed-up on the 
decorrelation of $\mathbf p$ because the proposals are uniform on all angles 
and the $\theta_j \to \theta_j+\pi$ transform is always accepted. 

\end{document}